\documentclass[a4paper,fleqn,usenatbib]{mnras}

\usepackage[T1]{fontenc}
\usepackage{ae,aecompl}
\usepackage{gensymb}
\usepackage{graphicx}	
\usepackage{amsmath}	
\usepackage{amssymb}	

\usepackage{newtxtext,newtxmath}

\title[The superluminous SN~2019neq]{Detailed spectrophotometric analysis of the superluminous and fast evolving SN 2019neq}

\author[A. Fiore et al.]{Achille Fiore,$^{1,2,3}$\thanks{E-mail: achillefiore@gmail.com}
Stefano Benetti,$^{3}$
Leonardo Tartaglia,$^{4,3}$
Anders Jerkstrand,$^{5}$
Irene Salmaso,$^{3,6}$
\newauthor
Lina Tomasella,$^3$
Antonia Morales-Garoffolo,$^{7}$
Stefan Geier,$^{8,9}$
Nancy Elias-Rosa,$^{3,10}$
Enrico Cappellaro,$^3$
\newauthor
Xiaofeng Wang,$^{11,12}$
Jun Mo,$^{11}$
Zhihao Chen,$^{11}$
Shengyu Yan,$^{11}$
Andrea Pastorello,$^3$
Paolo A. Mazzali,$^{13,14}$
\newauthor
Riccardo Ciolfi,$^{3,15}$
Yongzhi Cai,$^{16,17,18}$
Morgan Fraser,$^{19}$
Claudia P. Guti\'errez,$^{20,21}$
\newauthor
Emir Karamehmetoglu,$^{22,5}$
Hanindyo Kuncarayakti,$^{23}$
Shane Moran,$^{23}$
Paolo Ochner,$^{3,6}$
\newauthor
Andrea Reguitti,$^{24,3}$
Thomas M. Reynolds,$^{25,23}$
Giorgio Valerin$^{3,6}$
\\
$^{1}$Institut f\"ur Theoretische Physik, Goethe Universit\"at, Max-von-Laue-Str. 1, 60438 Frankfurt am Main, Germany.\\
$^{2}$INFN-TIFPA, Trento Institute for Fundamental Physics and Applications, Via Sommarive 14, I-38123 Trento, Italy.\\
$^{3}$INAF - Osservatorio Astronomico di Padova, Vicolo dell’Osservatorio 5, I-35122 Padova, Italy.\\
$^{4}$INAF - Osservatorio Astronomico d’Abruzzo, via M. Maggini snc, I-64100 Teramo, Italy\\
$^{5}$The Oskar Klein Centre, Department of Astronomy, Stockholm University, AlbaNova, SE-10691 Stockholm, Sweden.\\
$^{6}$Dipartimento di Fisica e Astronomia ‘G. Galilei’, Universit\`a di Padova, Vicolo dell’Osservatorio 3, I-35122 Padova, Italy.\\
$^{7}$Department of Applied Physics, School of Engineering, University of Cádiz, Campus of Puerto Real, E-11519 Cádiz, Spain.\\
$^{8}$Instituto de Astrofisíca de Canarias, 38200 La Laguna,Tenerife, Canary Islands, Spain.\\
$^{9}$GRANTECAN: Cuesta de San José s/n, 38712 Breña Baja, La Palma, Spain.\\
$^{10}$Institute of Space Sciences (ICE, CSIC), Campus UAB, Carrer de Can Magrans s/n, 08193 Barcelona, Spain.\\
$^{11}$Physics Department and Tsinghua Center for Astrophysics (THCA), Tsinghua University, Beijing, 100084, China.\\
$^{12}$Beijing Planetarium, Beijing Academy of Science and Technology, Beijing 100044, China.\\
$^{13}$Astrophysics Research Institute, Liverpool John Moores University, Liverpool, L3 5RF, UK.\\
$^{14}$Max-Planck Institute for Astrophysics, Garching, Germany.\\
$^{15}$INFN – Sezione di Padova, Via Francesco Marzolo 8, I-35131 Padova, Italy.\\
$^{16}$Yunnan Observatories, Chinese Academy of Sciences, Kunming 650216, PR China.\\
$^{17}$Key Laboratory for the Structure and Evolution of Celestial Objects, Chinese Academy of Sciences, Kunming 650216, PR China.\\
$^{18}$International Centre of Supernovae, Yunnan Key Laboratory, Kunming 650216, P.R. China\\
$^{19}$School of Physics, O'Brien Centre for Science North, University College Dublin, Belfield, Dublin 4, Ireland.\\
$^{20}$Institut d’Estudis Espacials de Catalunya (IEEC), Gran Capit\`a, 2-4,
Edifici Nexus, Desp. 201, E-08034 Barcelona, Spain.\\
$^{21}$Institute of Space Sciences (ICE, CSIC), Campus UAB, Carrer de Can
Magrans, s/n, E-08193 Barcelona, Spain.\\
$^{22}$Department of Physics and Astronomy, Aarhus University, Ny Munkegade 120, DK-8000 Aarhus C, Denmark.\\
$^{23}$Department of Physics and Astronomy, University of Turku, Vesilinnantie 5, 20500 Finland.\\
$^{24}$INAF - Osservatorio Astronomico di Brera, Via Bianchi 46, 23807 Merate (LC), Italy.\\
$^{25}$Cosmic DAWN Centre, Niels Bohr Institute, University of Copenhagen, R\aa dmandsgade, 62-64, 2200, Copenhagen, Denmark.
}

\date{Accepted XXX. Received YYY; in original form ZZZ}

\pubyear{2022}

\begin{document}
\label{firstpage}
\pagerange{\pageref{firstpage}--\pageref{lastpage}}
\maketitle

\begin{abstract}
SN~2019neq was a very fast evolving superluminous supernova. At a redshift $z=0.1059$, its peak absolute magnitude was ${-21.5\pm0.2}$ mag in $g$ band. In this work, we present data and analysis from an extensive spectrophotometric follow-up campaign using multiple observational facilities. Thanks to a nebular spectrum of SN~2019neq, we investigated some of the properties of the host galaxy at the location of SN~2019neq and found that its metallicity and specific star formation rate are in a good agreement with those usually measured for SLSNe-I hosts. We then discuss the plausibility of the magnetar and the circumstellar interaction scenarios to explain the observed light curves, and interpret a nebular spectrum of SN~2019neq using published \textsc{sumo} radiative-transfer models. The results of our analysis suggest that the spindown radiation of a millisecond magnetar with a magnetic field $B\simeq6\times10^{14}\,\mathrm{G}$ could boost the luminosity of SN~2019neq.
\end{abstract}




\section{Introduction}
\label{sec:intro}
It is widely accepted that supernovae (SNe) are a possible final stage of the life of massive stars and white dwarfs in close binary systems; they are observationally classified as type-I (SNe I) or type-II (SNe II) depending on whether they are hydrogen-poor or hydrogen-rich, respectively. SNe are usually discovered by untargeted wide-field surveys and they can be identified in their local environment performing differential photometry \citep[see also Sec.~\ref{sec:photometry} and][]{kessleretal2015}: in fact, their light curves (LCs) have a magnitude at maximum spanning the range $-14$ to $-19$ mag in the optical bands. SN light curves are normally interpreted as being powered by $^{56}\mathrm{Ni}$ radioactive decay \citep[e.~g.][]{nadyozhin1994} and by the energy deposited in the ejecta by the shock break-out. However, observations of the so-called superluminous SNe (SLSNe) have complicated this picture, as their absolute magnitude at maximum can be even brighter than $-21$ mag in optical bands \citep[e.g.][]{galyam2012,galyam2019}. Similar to their lower-luminosity siblings, these events are grouped in SLSNe I and SLSNe II based on the strength of their H features, however their luminosity cannot be comfortably explained by the energy released from $^{56}$Ni decay if the classical neutrino-wind driven core-collapse scenario is assumed. In fact, the explosion of standard SN progenitors synthesizes $\sim0.1\,\mathrm{M}_\odot$, while SLSNe require $\gtrsim1-10\,\mathrm{M}_\odot$ of $^{56}\mathrm{Ni}$ \citep[e.~g.][]{umedaandnomoto2008,galyametal2009,kasenetal2011,dessartetal2012}. In principle, the explosion of a pair-instability SN \citep[e.~g.][]{yoshidaetal2016} is a channel to synthesize much more $^{56}\mathrm{Ni}$ than standard core-collapse SNe, but this scenario is disfavoured by the observed spectra of SLSNe \citep[see e.g.][]{kozyrevaetal2015,moriyaetal2019,mazzalietal2019}. Two main alternatives have been considered to reproduce the observational properties of SLSNe. The first requires a shock where the SN ejecta collides with circumstellar material (CSM) lost by the progenitor star before its explosion \citep{woosleyetal2007,sorokinaetal2016,tolstov2017,woosley2017}. In this scenario, the shock driven by their SN ejecta converts the kinetic energy in radiation via collisional excitation and ionisation processes. Alternatively, the deposition in the ejecta of the spindown radiation of a newly-born, highly-magnetized neutron star \citep[a magnetar, e.g.][]{woosleyetal2007,kasenandbildsten2010} can boost SLSNe luminosities. However, the magnetar model does not naturally predict the bumps often seen in SLSNe-I LCs both before and after maximum \citep[see e.g.][but see \citealp{moriyaetal2022}]{nicholletal2016,smithetal2016,vreeswijketal2017,nicholletal2017,lunnanetal2020,fioreetal2021,gutierrezetal2022,chenetal2023a,chenetal2023b,westetal2023,linetal2023} but is able to account for the huge variety of LC evolutionary timescales shown by SLSNe I \citep{inserraetal2013,chatzopoulosetal2013,nicholletal2014,nicholletal2015,nicholletal2017}.

Observationally, SLSNe I have blue spectra (with a blackbody temperature spanning $T_{\rm BB}=10000-20000$ K) during the pre-maximum/maximum phases, with prominent absorptions between 3000-5000 \AA{}. These features are usually identified as \ion{O}{ii} \citep[e.~g.][]{mazzalietal2016}, although there is no general consensus on this (as discussed later). After 2-3 weeks from maximum, SLSNe I enter a new phase in which their spectra bear a resemblance to those of broad lined SNe Ic (SNe Ic BL) at their maximum luminosity \citep[e.g. ][]{pastorelloetal2010}. Photometrically, their LCs are very heterogeneous and may evolve over a wide range of timescales \citep[see e.g. Fig.~5 of][]{deciaetal2018}. \citet{inserraetal2018b} suggested to distinctly separate rapidly declining SLSNe I from slow ones, but larger SLSNe-I samples point towards a continuous distribution of timescales \citep{nicholletal2015,deciaetal2018,lunnanetal2018b,angusetal2019}. Recently, \citet{koenivestothandvinko2021} and \citet{koenyvestoth2022} proposed a new SLSNe-I subclassification based on their features in pre-maximum/maximum spectra: type-W SLSNe I, whose absorptions are well fitted by \ion{O}{ii} at reasonable velocities and type-15bn SLSNe I, whose early spectral features are not easily explained by \ion{O}{ii} and show similarities with the coeval spectra of SN 2015bn \citep{nicholletal2016}. 

In the present work, we deal with the SLSN I SN~2019neq, which was classified by \citet{koenivestothandvinko2021} in the type-W subgroup. At a redshift $z\simeq0.1059$ (see Sec.~\ref{sec:spectroscopy}), SN~2019neq is located at ${\rm RA}= 17^{\rm h}54^{\rm m}26\fs736$, ${\rm Dec}=+47\degr15\arcmin40\farcs62$ and it is likely associated with the galaxy SDSS J175426.70+471542.3. It was discovered on 2019 August 9 by the Zwicky Transient Facility \citep[ZTF, ][]{bellmetal2019} with an apparent magnitude of $g=20.4$ mag \citep{perleyetal2019}, and named with the internal designation of ZTF19abpbopt. A few days later, a spectrum observed with the Palomar 60-inch+SED Machine revealed a hot continuum with some unidentified features \citep{perleyetal2019}. After $\sim18$ days, a spectrum of SN~2019neq was taken at the {Liverpool Telescope} \citep[LT, ][]{steeleetal2004} (Roque de Los Muchachos Observatory, La Palma, Spain) equipped with the SPRAT \citep[SPectrograph for the Rapid Acquisition of Transients, ][]{piasciketal2014} instrument and classified as a SLSN~I \citep{koenivestothetal2019}. A $g=17.2$ mag detection implied that it was still in the rising phase with a rest-frame rate of $\sim0.2$ mag/day. 

SN 2019neq was already included in several studies: \citet{konyvestothetal2020} presented and discussed the ZTF photometry of SN 2019neq and three photospheric spectra. Furthermore, SN 2019neq was included in four sample papers \citep{hosseinzadehetal2022,chenetal2023a,chenetal2023b,pursianenetal2023}: in particular,  \citet{hosseinzadehetal2022} and \citet{chenetal2023a,chenetal2023b} interpret post-maximum undulations in the $g$- and $r$-filter LCs of SN 2019neq as SLSNe-I bumps, i. e. they attributed the undulations either to a variable energy source or to an external one (e.g. CSM interaction). \citet{pursianenetal2023} analyzed linear-polarimetry data of 7 H-poor SLSNe I and concluded that SLSNe I with oscillating LCs usually show an increase of the degree of polarimetry: therein, SN 2019neq was included in the non-oscillating SLSNe I subsample. In this work, we present a deep photometric and spectroscopic dataset of SN 2019neq and, based on new coeval photometry, disfavour that the undulations seen in its $g$- and $r$-filters LCs have an astrophysical origin. Our analysis is in agreement with the ejecta mass estimated by \citet{koenivestothetal2020} in an independent way.

In detail, in Sec.~\ref{sec:photometry} and \ref{sec:spectroscopy} we introduce the  photometric and the spectroscopic observations of SN~2019neq, respectively; in Sec.~\ref{sec:discussion} we compare SN~2019neq LCs and spectra with other SLSNe I (Sec. \ref{sec:comparisons}), we study its color and temperature evolution (Sec.~\ref{sec:bbtemp}), and the evolution of its expansion velocity via the \ion{O}{ii} spectral absorption features (Sec.~\ref{sec:19neq_phvel}). Finally, we discuss the magnetar and the CSM-interaction interpretations (Sec.~\ref{sec:19neq_model}) using the nebular spectrum of SN~2019neq (Sec.~\ref{sec:19neq_sumo}) {and} modelling its observed multicolor LCs (Sec.~\ref{sec:mosfit}).
Throughout the paper, we will assume a flat Universe with $\Omega_\mathrm{M}=0.31$, $\Omega_\Lambda=0.69$ and with a Hubble constant $H_0=71\pm3\,\mathrm{km\,s^{-1}\,Mpc^{-1}}$; we took this value of $H_0$ as an average among the estimates provided by {\citet{planck2016}, \citet{kethanetal2021} and \citet{riessetal2021}}. 

\section{Photometry}
\label{sec:photometry}
\subsection{Observations and data reductions}
Ultraviolet/optical/near-infrared imaging data of SN~2019neq were obtained via different facilities. In detail, we used the NOT Unbiased Transients Survey 2 \citep[NUTS2\footnote{\texttt{https://nuts.sn.ie}}, ][]{mattilaetal2016,holmboetal2019} at the 2.56-m Nordic Optical Telescope (NOT)+{ALFOSC}/NOTCam, the 2.0-m Liverpool Telescope (LT)+IO:O, La Palma observatory, Spain; the 1.82-m Copernico Telescope+AFOSC and Schmidt telescopes at the Asiago Astrophysical Observatory, Italy; the Tsinghua-NAOC (National Astronomical Observatories of China) Telescope (TNT)+BFOSC (Beijing Faint Object Spectrograph and Camera), Xinglong Observatory, China \citep{wangetal2008,huangetal2015}. $uvw2,uvm2,uvw1,U,B,V$ follow up was triggered with the \textit{Neil Gehrels Swift} Observatory+Ultraviolet/Optical Telescope \citep[UVOT, ][]{gehrelsetal2004}. \textit{Swift}/UVOT $uvw2,uvm2,uvw1,U,B,V$-filter frames were reduced\footnote{Their calibration was done using the updated version (November 2020) of the sensitivity corrections.} with the \textsc{heasoft} package \citep[version 6.25][]{Heasoft2014a}. The ground-based $u,B,V,g,r,i,z$ photometric frames were preliminary pre-processed, i.e. they were corrected for overscan, bias and flat field. Magnitude measurements were performed by means of the \textsc{ecsnoopy} pacakage\footnote{\textsc{ecsnoopy} is a python package for SN photometry using PSF fitting and/or template subtraction developed by E. Cappellaro. A package description can be found at \texttt{http://sngroup.oapd.inaf.it/} .} \citep{cappellaro2014}. Before measuring the SN flux, we accounted for the contribution of the background {contaminating the SN light} either with a polynomial interpolation or with the template-subtraction technique. The latter was performed with the \textsc{hotpants} tool \citep{becker2015}. When possible, we used template frames obtained with the same instrumental setting used for the scientific observations and observed after that the SN faded well below the detection limit. When deep template frames were not available, we estimated the background contribution interpolating a low-order polynomial to the area surrounding the SN position. This was interpreted as the background level and then subtracted from the photometric frames contaminated by the SN background. SN magnitudes were then measured fitting a point spread function to the SN \citep{stetson1987}. A detailed description of the image reduction and measurements procedures can be found in \citet{fioreetal2021}. Instrumental $u,B,g,V,r,i,z$ magnitudes were calibrated on a sequence of non-saturated field stars from the SDSS (Sloan Digital Sky Survey) and the Pan-STARRS \citep[Panoramic Survey Telescope and Rapid Response System, ][]{chambersetal2016} surveys, respectively. For $U,B,V$ filters, we converted the reference star magnitudes from Sloan to Johnson system following \citet{chonisandgaskell2008}. $J,H,K_{\rm s}$ magnitudes were calibrated with reference to field stars present in the 2MASS catalogue \citep[Two-Micron All Sky Survey,][]{skrutskieetal2006}. Given the very rapid evolution of SN~2019neq, deep $u,B,V,g,r,i,z$ template frames were observed on 2020 June 12 ($\mathrm{MJD}=59012$, only 281 rest-frame days after the maximum luminosity) by NUTS2. Since the SN was not detected in the template frames within a 3$\sigma$ detection limit, we assumed that the luminosity of SN~2019neq faded well below the galaxy luminosity. Finally, it was not possible to template subtract the $J,H,K_{\rm s}$-filter frames since no suitable template was available for these filters. Therefore we estimated the background contribution fitting a low-order polynomial. We also include $o$ and $c$ ATLAS \citep[Asteroid Terrestrial-impact Last Alert System, ][]{tonryetal2018} photometry, which we converted to standard $g$ and $r$ via \citet{tonryetal2018a}. The results of our magnitude measurements are listed in Tabs.~\ref{tab:19neq_uvottab}, \ref{tab:19neq_ugriztab}, \ref{tab:19neq_bvtab}, \ref{tab:19neq_jhktab}. Finally, we applied to these data the S- and K- corrections to bring back the many instrumental photometric systems to the corresponding standard and rest-frame ones (see Appendix \ref{sec:skcorrs} for details).
\subsection{Observed and bolometric light curves}
\label{sec:obslcs}
\begin{figure*}
    \centering
    \includegraphics[width=\textwidth]{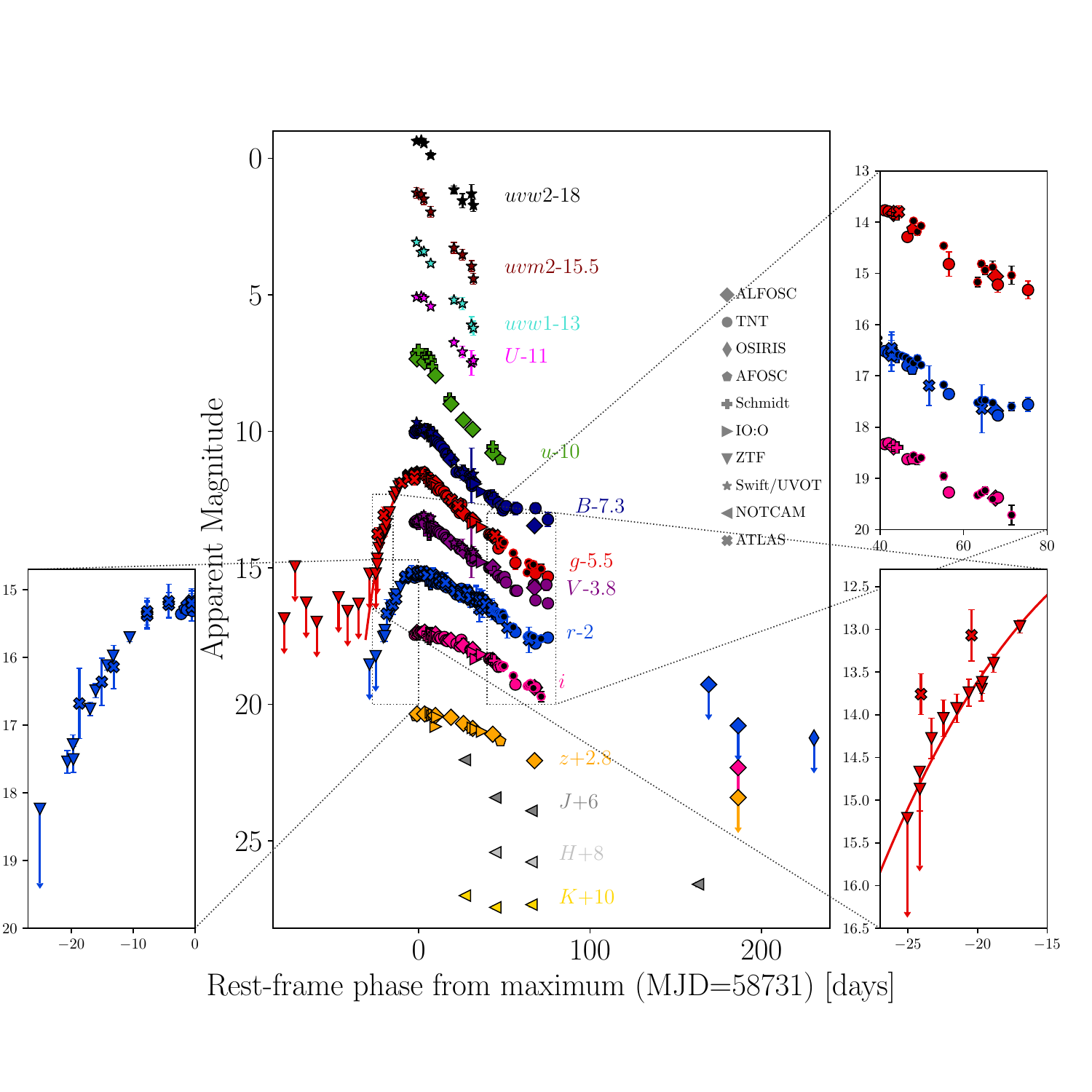}
    \caption{S-corrected lightcurves for SN 2019neq in $uvw2, uvm2, uvw1, U, u, B, g, V, r, i, z, J, H, K_{\rm s}$ bands plotted in black, brown, cyan, pink, green, dark blue, red, purple, blue, magenta, orange, dark silver, silver, and yellow, respectively. Photometry obtained with different instruments are plotted with different symbols, as labelled in the upper right-hand corner. The $g$-filter LC was fitted with a high-order polynomial (red solid line) to obtain the epoch of maximum luminosity (see text). Arrows correspond to 2.5$\sigma$ detection limits. Magnitudes are in the AB system. The upper inset on the right-hand side shows the period in which the late bump is visible (additional data from \citet{hosseinzadehetal2022} are represented by black dots); the lower-left and -right insets show the pre-maximum phases in $r$ and $g$ filters, respectively}.
    \label{fig:lcs}
\end{figure*}

The S-corrected LCs are shown in Fig.~\ref{fig:lcs}. To investigate the presence of a possible pre-maximum bump feature in the LC, and to better constrain the rising phase towards the maximum luminosity, we added publicly-available ZTF $g$- and $r$-filter photometry. The latest non-detection fainter than the polynomial fit of the early $g$ LC corresponds to 2019 August 7 ($\mathrm{MJD} = 58703.312$, $g=20.7$ mag), and the first detection in $g$ filter ($g=20.4$ mag) was about one day later ($\mathrm{MJD}=58704.307$). It is therefore possible to estimate that the explosion date occurred on $\mathrm{MJD}\simeq58704\pm1$ (2019 August 9). Early $g$-filter detection limits exclude the occurrence of a pre-maximum bump within $\sim79$ days before the estimated explosion date. To estimate the maximum-luminosity epoch, we fitted a high-order polynomial to the $g$-filter LC and find that the maximum occurred on $\mathrm{MJD}_{\rm MAX}=58731\pm2$ (2019 September 5) at a magnitude $g_{\rm MAX}=17.07\pm0.10$ mag (uncertainties were established by varying the order of the fitted polynomial). Given a luminosity distance $d_{\rm L}=482.2^{+21.3}_{-19.5}$ Mpc, corresponding to a distance modulus $\mu=38.416^{+0.094}_{-0.090}$ mag, and a Galactic extinction $A_{V}=0.104$ mag \citep{schlaflyetal2011}, the peak absolute magnitude of SN~2019neq is ${M_g=-21.5\pm0.2}$ mag. As no narrow absorption lines from the \ion{Na}{I}D doublet \citep{poznanskietal2012} are seen in the spectra, we assume no extinction from the host galaxy. In the following, we refer to the rest-frame time with respect to maximum as ``phase'' ($\phi$). The assumed explosion epoch implies that the phase at the point of explosion is $-24$ days. 

\begin{figure}
    \centering
    \includegraphics[width=0.45\textwidth]{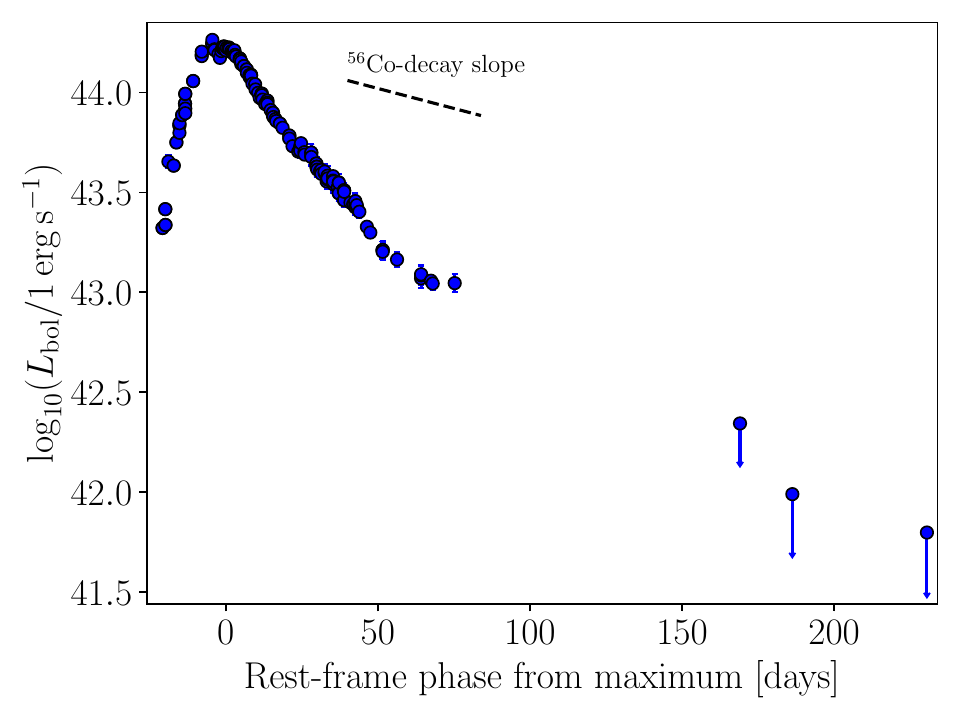}
    \caption{Pseudo-bolometric LC of SN~2019neq. For comparison also the $^{56}\mathrm{Co}$-decay slope is also shown (black-dashed line).}
    \label{fig:bollc}
\end{figure}

The pseudo-bolometric LC of SN~2019neq was computed by integrating the broad band photometry, adopting as reference the epochs of the $r$-filter photometry (see Fig.~\ref{fig:bollc}, Tab.~\ref{tab:19neq_kcorr} for the K-correction and Tab.~\ref{tab:19neq_blc} for the pseudo-bolometric LC). SN2019neq rises towards the maximum luminosity very rapidly at a rate $v_{\rm rise}\approx-0.13\,\mathrm{mag}\,\mathrm{d}^{-1}$, it reaches a maximum luminosity of $(2.04\pm0.04)\times10^{44}\,\mathrm{erg\,s^{-1}}$ and fades at a rate $v_{\rm decline}\approx0.05\,\mathrm{mag}\,\mathrm{d}^{-1}$, which is $\sim2.6$ times slower than the rising rate. This ratio is somewhat higher than the typical $v_{\rm decline}/v_{\rm rise}\approx2$ ratio for SLSNe~I \citep{nicholletal2015}.
\subsection{LC bumps in SN 2019neq}
\label{sec:lcbumps}
LCs of SLSNe I often display pre-maximum and/or post-maximum bumps, which are usually interpreted as signatures of CSM-interaction. Their occurrence seems to be more common for the slower-evolving events \citep[see e.~g.][]{yanetal2015,yanetal2017,nicholletal2015b,nicholletal2016b}, but this connection is debated and might simply be due to the fact that slow SLSNe I can be observed for a longer time.

\citet{hosseinzadehetal2022} and \citet{chenetal2023b} noticed post- and pre-maximum undulations in the SN 2019neq LCs respectively, and included it in sample studies of bumpy SLSNe I. In particular, \citet{hosseinzadehetal2022} used ZTF, Pan-STARRS and ATLAS data, together with $g,r,i$ observations from the 1.2m telescope+KeplerCam at the Fred Lawrence Whipple Observatory for SN~2019neq \citep{szentgyorgyietal2005}. Their bumpy-SLSNe I sample consists of 34 objects whose LCs deviate from the best-fit magnetar-powered \textsc{mosfit} LC, which introduces a model dependency in the definition of a `bump'. Their KeplerCam $g,r,i$ measurements are in good agreement with our coeval photometric data, which indeed seem to show a flattening at about $~80\,\mathrm{days}$ after maximum (see Fig.~\ref{fig:lcs}, upper-right panel) and reveal $\lesssim0.3\,\mathrm{mag}$-amplitude undulations occurring on a timescale of $\lesssim2\,\mathrm{days}$. However, given the interruption of the follow-up at this phase, we are not able to reach a firm conclusion on the veracity of physical origin of these LC features.

\citet{chenetal2023b} discussed ZTF photometry which we also use in this work. In the pre-maximum data (at a phase of $\sim-20$), the $g$ and possibly $r$ LCs appear to display a hump (see Fig.~\ref{fig:lcs}, lower-right panel). This deviation from the general rising slope of the LCs is however encompassed by the errorbars, although the possible correspondence two bands (the only in which we have very early detections) might provide some additional support. However, as evidence for the presence of bumps is relatively weak, we do not consider data from \citet{hosseinzadehetal2022} in the following analysis.

\section{Spectroscopy}
\label{sec:spectroscopy}
\subsection{Observations and data reductions}
Optical spectra of SN~2019neq were obtained with the 1.82m Copernico+AFOSC telescope, with NOT+ALFOSC via NUTS2, with the Xinglong 2.16-m telescope+BFOSC and with the 10.4m Gran Telescopio CANARIAS (GTC)+OSIRIS \citep[Optical System for Imaging and low-Intermediate-Resolution Integrated Spectroscopy,][]{cepaetal2000}. The raw two-dimensional spectroscopic frames were preprocessed, wavelength-calibrated, extracted and flux-calibrated with standard \textsc{iraf} procedures called via the graphical user interface\footnote{\texttt{http://sngroup.oapd.inaf.it/} .} \textsc{foscgui}\footnote{\textsc{foscgui} is a \textsc{python/pyraf} based graphic user interface aimed at extracting
SN spectroscopy and photometry obtained with FOSC-like instruments.
It was developed by E. Cappellaro. A package description
can be found at \texttt{http://sngroup.oapd.inaf.it/foscgui.html} .}. BFOSC spectra were reduced with \textsc{iraf} procedures directly. The flux calibration of each spectrum was then checked against the coeval optical photometry (which was interpolated in case of missing epochs).
\subsection{The spectra}
\begin{figure*}
    \centering
    \includegraphics[width=0.8\textwidth]{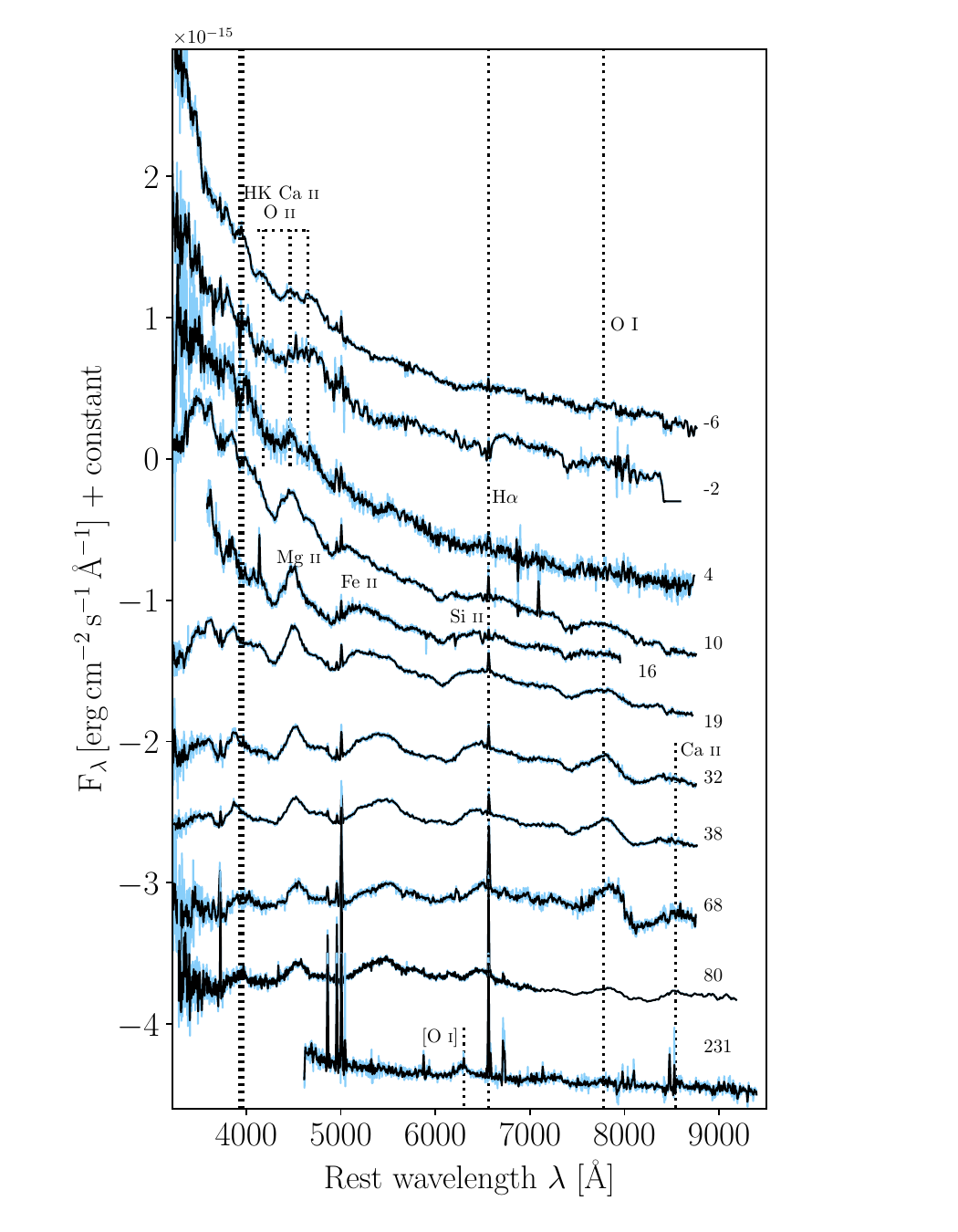}
    \caption{The spectral evolution of SN~2019neq reported to the rest-frame. The spectra (light blue solid lines) were also smoothed with a Savitzky–Golay filter \citep[][black solid lines]{savitzkyandgolay1964} to make the features more distinguishable. Spectra were offset and scaled. Identified spectral features are marked with black dashed lines at their corresponding rest-frame wavelength. Numbers on the right-hand side of each spectrum indicate their phase. }
    \label{fig:19neq_spec}
\end{figure*}

The spectral evolution of SN~2019neq is shown in Fig.~\ref{fig:19neq_spec} (see also Tab. \ref{tab:19neq_sfo}). Pre-maximum/maximum spectra of SN~2019neq have a hot blue continuum with black-body temperatures of about $T_\mathrm{BB}\approx16000\,\mathrm{K}$. The hot continuum is almost featureless, with only a W-shaped \ion{O}{ii} absorption feature and the \ion{O}{I} $\lambda\,7774$ feature visible in the red part. At bluer wavelengths, the \ion{Ca}{II} H\&K doublet is also present.
After 10 days from maximum, the continuum cools, the W-shaped features gradually disappear in favour of the \ion{Fe}{II}  and \ion{Mg}{II} features and the spectra of SN~2019neq start to resemble a SN Ic BL spectrum at maximum luminosity. {In addition}, at this phase \ion{Si}{II} was identified by \citet{konyvestothetal2020} with a spectral fit obtained with \textsc{syn++}.
After +19 days, the continuum gradually cools down, and up to +80 days the spectral features do not evolve significantly, except for their intensity and velocity (see Sec.~\ref{sec:19neq_phvel}). At late phases, when the forbidden emission lines start to be seen in the spectrum, the \ion{Mg}{ii} feature is expected to be contaminated or replaced by the semi-forbidden \ion{Mg}{i}] $\lambda\,4571$ feature; however, this region is not covered by the latest spectrum (+231 days). Finally, we used the positions of the narrow H$\alpha$, [\ion{O}{iii}] host-galaxy emission lines measured in the +68, +80 and +231 days spectra to derive a
host galaxy redshift of $z=0.10592\pm0.00005$ (where the uncertainty is given by the dispersion of the measurements).

We also estimated the metallicity at the site of SN~2019neq by means of the narrow emission lines from the host galaxy which show up at late/nebular phases. We measured the flux emitted within these lines after optimizing the extraction from the two-dimensional spectrum for the host-galaxy and calibrating it against the host photometry at the SN location \citep[see also][]{lunnanetal2014}. To derive the metallicity, we used the tool \textsc{pymcz} \citep{biancoetal2016}. In short, \textsc{pymcz} computes the metallicity from the flux measurements of those narrow emission lines via a number of metallicity diagnostics. To associate the errorbars to the metallicty measurements, \textsc{pymcz} randomly samples a Gaussian distribution whose mean and standard deviation are given by the flux measurements and their uncertainties, respectively. In the case of SN~2019neq, we measured the flux emitted within the [\ion{O}{II} ]$\lambda\,3727$, H$\beta$, [\ion{O}{III}] $\lambda\,4959$, [\ion{O}{III}] $\lambda\,5007$, H$\alpha$, [\ion{S}{II}] $\lambda\,6717$ emission lines. The nebular spectrum of SN~2019neq at 231 days after maximum does not show the [\ion{N}{II}] $\lambda\lambda\,6548,6583$ narrow emission lines. The only diagnostic which could be suitably used is then the \textsc{m08\_o3o2} \citep{maiolinoetal2008}, giving a metallicity $12+\log_{10}(\mathrm{O/H})\simeq8.3$, which corresponds to $Z\simeq0.4\,\mathrm{Z}_\odot$ assuming a solar metallicity $12+\log_{10}(\mathrm{O/H})=8.69$ \citep{asplundetal2009}.

Finally, we estimated the star formation rate (SFR) of the host galaxy of SN~2019neq based on the measurements of the flux emitted by the reddening corrected narrow H$\alpha$ using equation 2 of \citet{kennicutt1998}. The derived SFR is $\simeq2.6\,\mathrm{M_\odot\,yr^{-1}}$ , similar to the SFRs measured by \citet{chenetal2017} for a sample of galaxies hosting SLSNe I. In addition, we estimated the specific SFR, i.e. the SFR expressed per unit of stellar mass content of the host galaxy, $\mathrm{sSFR}\equiv \mathrm{SFR}/M_{*}$. To estimate $M_{*}$, we used equation 8 of \citet{tayloretal2011} after having measured the $g_{\rm host}=20.4\pm0.1\,\mathrm{mag}$ and  $i_{\rm host}=19.7\pm0.1\,\mathrm{mag}$ for the host galaxy. We adopted Galactic extinction corrections from \citet{schlaflyetal2011} and we obtained K-corrections scaling a starburst-galaxy template from \citet{calzettietal1994} to the $g_\mathrm{host}$ and $i_{\rm host}$ magnitudes, thus obtaining $\log_{10}M_{*}/M_\odot\simeq9.1$ and $\mathrm{sSFR}\simeq2.3\,\mathrm{Gyr^{-1}}$. These values are in a good agreement with the average properties of a sample of 31 SLSNe I host galaxies studied by \citet{lunnanetal2014}. We stress that the SFR and sSFR values deduced can be highly affected by the error on the K-corrections.
\section{Discussion}
\label{sec:discussion}
\subsection{Photometric and spectroscopic comparisons}
\label{sec:comparisons}
\begin{figure*}
    \centering
    \includegraphics[width=0.8\textwidth]{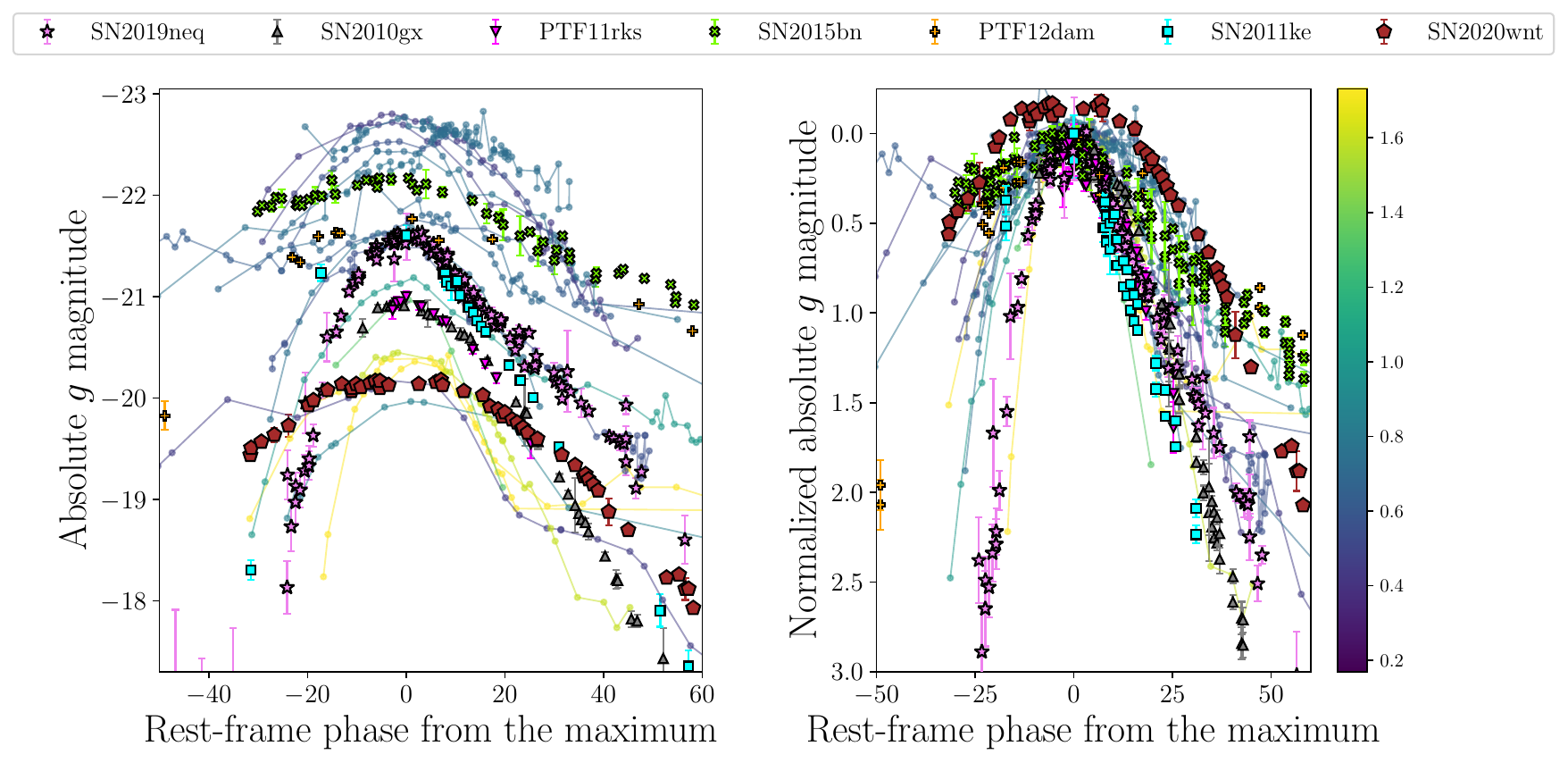}
    \caption{Comparisons of the absolute $g$-filter LC of SN~2019neq (black dots) with those of two fast-evolving SLSNe~I, SN~2011ke (cyan triangles) and PTF11rks (purple hexagons) \citep[data from][]{inserraetal2013} and two slow-evolving SLSNe~I, SN 2015bn (green diamonds) \citep{nicholletal2016} and PTF12dam (orange diamonds) \citep{nicholletal2013,chenetal2015,vreeswijketal2017}. Right panel: LCs comparison  normalized to the same maximum luminosity. Left panel: LCs comparison without rescaling. The absolute LCs were computed assuming the cosmological parameters used in this work (see Introduction).}
    \label{fig:19neq_phot_comp}
\end{figure*}
\begin{figure}
\centering
\includegraphics[width=0.45\textwidth]{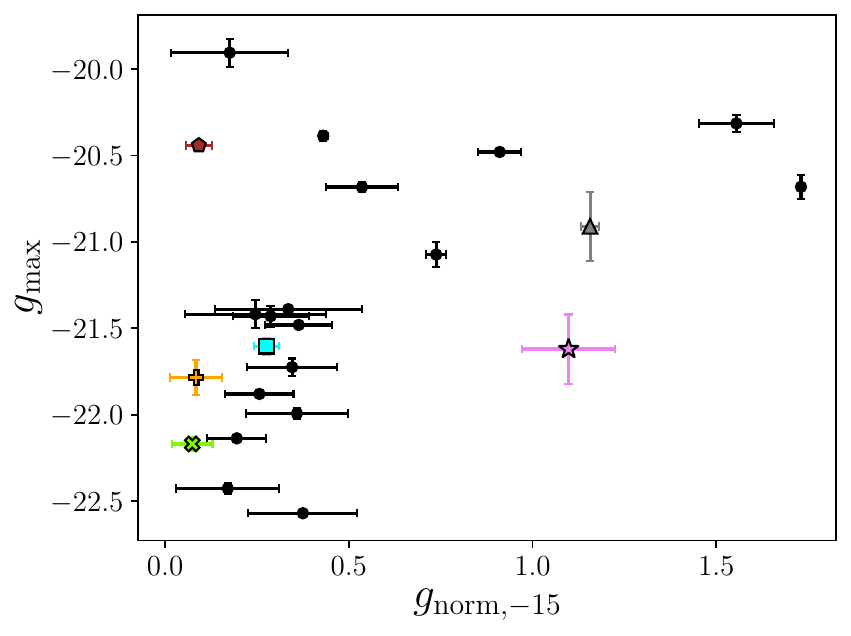}
\caption{Absolute $g$-filer magnitude $g_\mathrm{max}$ (corrected for foreground extinction and K-corrected) versus the normalized magnitude at $-15$ days $g_{\mathrm{norm}, -15}$ (see text). Black dots correspond to the ZTF sample (see Tab.~\ref{tab:ztfsample} for the coordinates), while the other symbols are the same as in Fig.~\ref{fig:19neq_phot_comp}.}
\label{fig:correlation}
\end{figure}
\begin{figure*}
    \centering
    \includegraphics[width=0.7\textwidth]{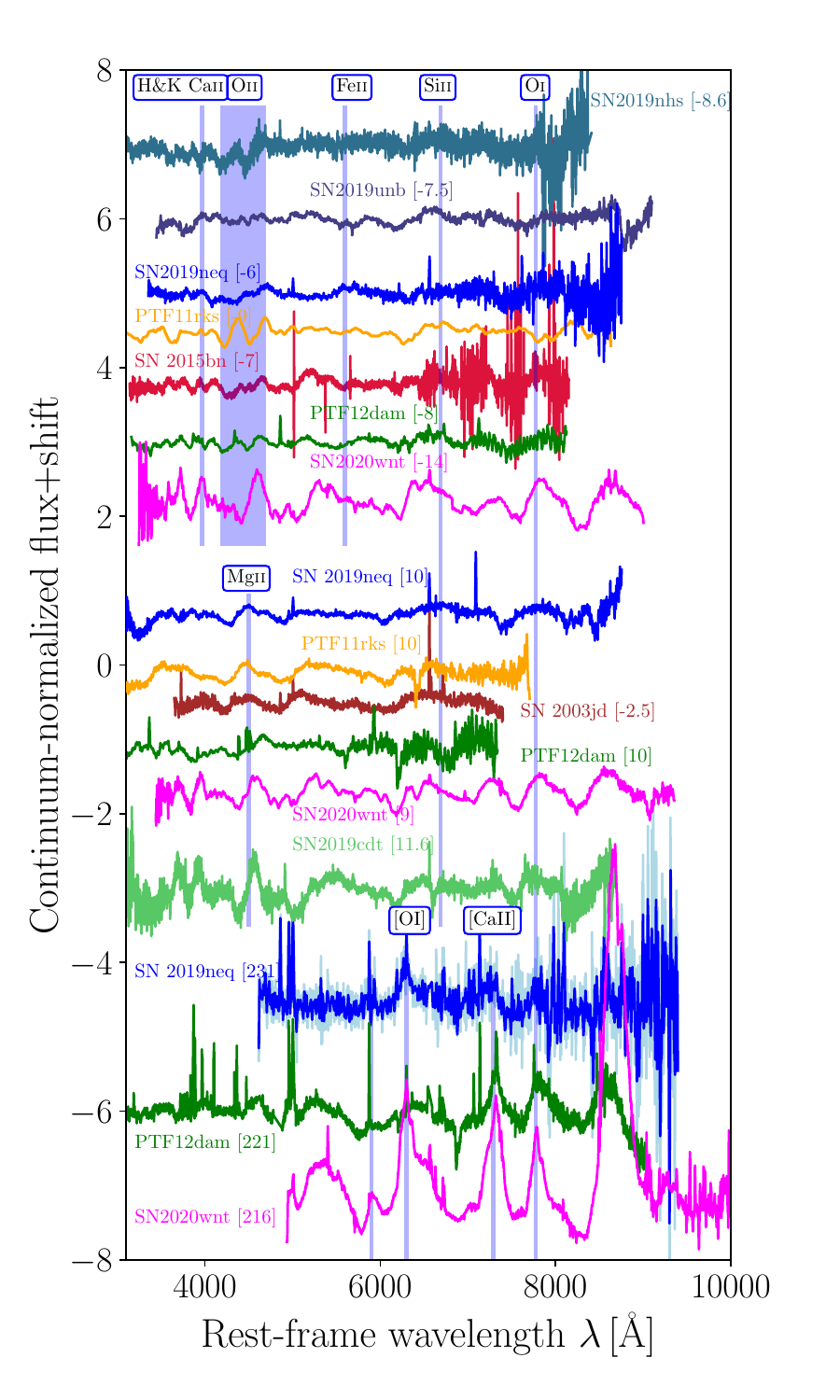}
    \caption{Comparison of three spectra of SN~2019neq (blue) with those of other SLSNe at broadly coeval phases with respect to maximum. The phases are reported in square brackets. Line identifications are marked with blue shaded areas and for each of them the corresponding ion is labelled on the top. For the last spectrum of SN 2019neq we plot both the original (light-blue) and its smoothed version (with a Savitzky-Golay filter, blue).}
    \label{fig:19neq_spec_comp}
\end{figure*}
In order to highlight its similarities/differences with other SLSNe I, we selected a sample including both very slow- and fast-evolving SLSNe~I to compare their K-corrected $g$-filter absolute LCs and spectra with those of SN~2019neq. The slow-SLSNe~I subsample consists of SN~2020wnt \citep{gutierrezetal2022}, SN~2015bn \citep{nicholletal2016} and PTF12dam \citep{nicholletal2013,chenetal2015,vreeswijketal2017}, while the fast-SLSNe~I one includes SN 2011ke and PTF11rks \citep{inserraetal2013,quimbyetal2018}. In addition, we imported the best-sampled and non-heavily oscillating $g$-filter ZTF LCs among those published in \citet{chenetal2023a} (see Tab.~\ref{tab:ztfsample}). We compared the LCs of the SLSNe I of this sample with that of SN 2019neq. In addition, we investigated if SN~2019neq has a peculiar behaviour compared to the expected correlation between the SLSNe I maximum luminosity and the evolutionary timescales  \citep[see][]{inserraetal2018b}.

The LC comparison was done both in absolute and normalized units (see Fig.~\ref{fig:19neq_phot_comp}, left and right panel, respectively). When the K-corrections were not available from the spectra, we approximate it as $-2.5\log_{10}{(1+z)}$ \citep{hoggetal2002}, which was shown to be a reasonable approximation for SLSNe I \citep{chenetal2023a}. The LCs of the ZTF sub-sample are coloured according to their evolutionary velocity: we used the normalized magnitude at a phase of -15 rest-frame days,  $g_{ \mathrm{norm},-15}$, as a proxy for it. Hence, faster-evolving SLSNe I have greater values of $g_{\mathrm{norm},-15}$ and vice versa. To measure $g_{\mathrm{norm},-15}$, for each SN we fitted a polynomial to the $g$-filter LC around the maximum luminosity. The degree of the polynomial and the fit domain were varied in order to minimize the root-mean squared of the fit. This quantity was also used as errorbar on $g_{\mathrm{norm}, -15}$, although this choice may overestimate it. The values of $g_{\mathrm{norm},-15}$ and $g_{\rm max}$ of the selected comparison sample are listed in Tab.~\ref{tab:ztfsample} too. We then plot the $g_{\rm max}$ values versus $g_{{\rm norm},-15}$ (Fig.~\ref{fig:correlation}). Our sample confirms the above-mentioned correlation between peak luminosities and the evolutionary timescales, with some exceptions nonetheless: SN 2019neq is actually one of the outliers as it evolves faster than the other SLSNe I with comparable peak luminosities; SN2018bym and the peculiar SN 2020wnt are slower than the other SLSNe I having comparable peak luminosities. On the contrary, SN 2015bn, PTF12dam, SN 2010gx and SN 2011ke better agree with the general trend  followed by the ZTF subsample (PTF11rks was excluded due to the lack of pre-maximum data). However, much larger samples are required to draw general conclusions on this.

Furthermore, we compared three spectra of SN~2019neq (at phases $-6$, +10, +231 days) with the available spectra for the comparison sample (including some of those belonging to the ZTF sample, see Fig.~\ref{fig:19neq_spec_comp}) at similar phases. In particular, we included pre-maximum spectra of SN 2019nhs and SN 2019unb and post maximum spectrum of SN 2019cdt. We also added a spectrum of the type-Ic BL SN 2003jd {at about its maximum luminosity given its resemblance to SLSNe~I spectra after maximum (see Sec.~\ref{sec:intro}) }. The absorptions on the blue part of the early spectrum of SN~2019neq are different from those of SN~2015bn, PTF11rks and SN~2020wnt, and hence due to other transitions \citep[see also][]{koenivestothandvinko2021}, while they possibly share the broad features from \ion{Si}{ii} and \ion{O}{ii} on the red side. In particular, the spectrum of SN 2019neq at $-6$ days looks more similar to the almost coeval one of SN~2019unb, whose location in the $(g_{\mathrm{norm}, -15}, g_{\rm max})$ space is also peculiar (see Tab.~\ref{tab:ztfsample}).
At later phases, the spectrum of SN~2019neq (+10 days after maximum) is similar to that of PTF11rks at the same phase and shows several broad features which nearly reproduce the spectral behaviour of SNe Ic BL at maximum luminosity. To see this, we compared the spectrum of SN~2019neq at 10 days after maximum with that of the SN~2003jd at maximum luminosity, which was given as best-match template by \textsc{gelato} \citep{harutyunyanetal2008}. Also the post-maximum spectrum of SN~2020wnt, PTF12dam and SN~2019cdt are similar to the other spectra shown at comparable phases, but the prominent feature seen in these spectra at about 4700 \AA{} cannot be easily explained by the \ion{Mg}{ii} $\lambda$4571 only.

Finally, the spectrum of SN~2019neq observed $231$ days after maximum was compared with the spectra of PTF12dam and SN~2020wnt. At these epochs, the spectra look different. In particular, the nebular spectra of PTF12dam and SN~2020wnt still show some broad features between 5000 and 6000 \AA{} and more prominent [\ion{Ca}{ii}] $\lambda\lambda$ 7291, 7323 and \ion{O}{i} $\lambda$ 7774 features. The nebular spectrum of SN~2019neq looks instead pretty much featureless, except for the [\ion{O}{i}] $\lambda\lambda$ 6300,6364; however, its lower signal-to-noise ratio makes a more thorough comparison difficult.
\subsection{SED, colours, photospheric temperature and radius evolutions}
\label{sec:bbtemp}
\begin{figure}
    \centering
    \includegraphics[width=0.5\textwidth]{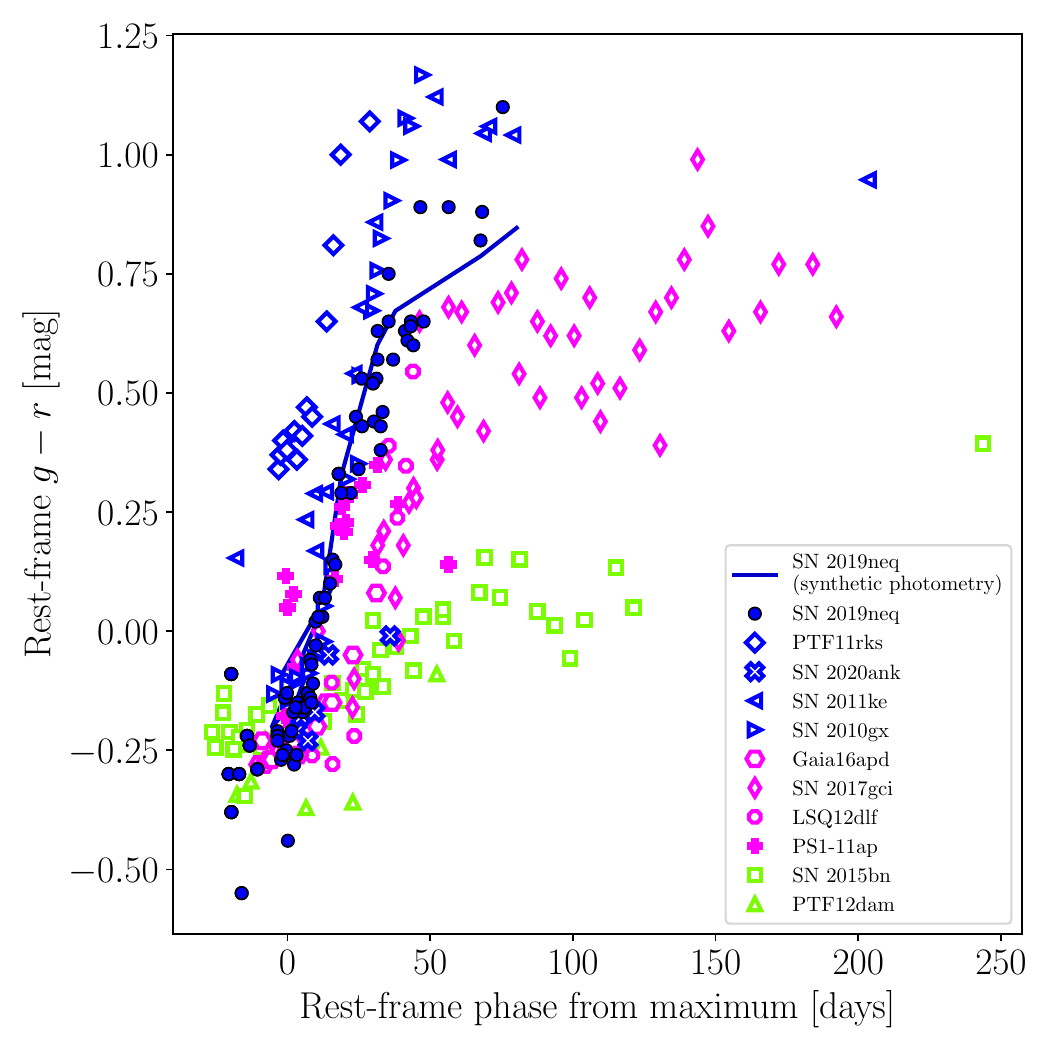}
    \caption{Rest-frame $g-r$ evolution of SN~2019neq (blue filled dots) compared to that of a sample of fast-evolving (blue-edged markers), intermediate (magenta-edged markers) and slow-evolving SLSNe I (green-edged markers). The solid-blue line joins the rest-frame $g-r$ points computed via synthetic photometry on the spectra of SN~2019neq.}
    \label{fig:gmr}
\end{figure}

\begin{figure*}
    \centering
    \includegraphics[width=0.85\textwidth]{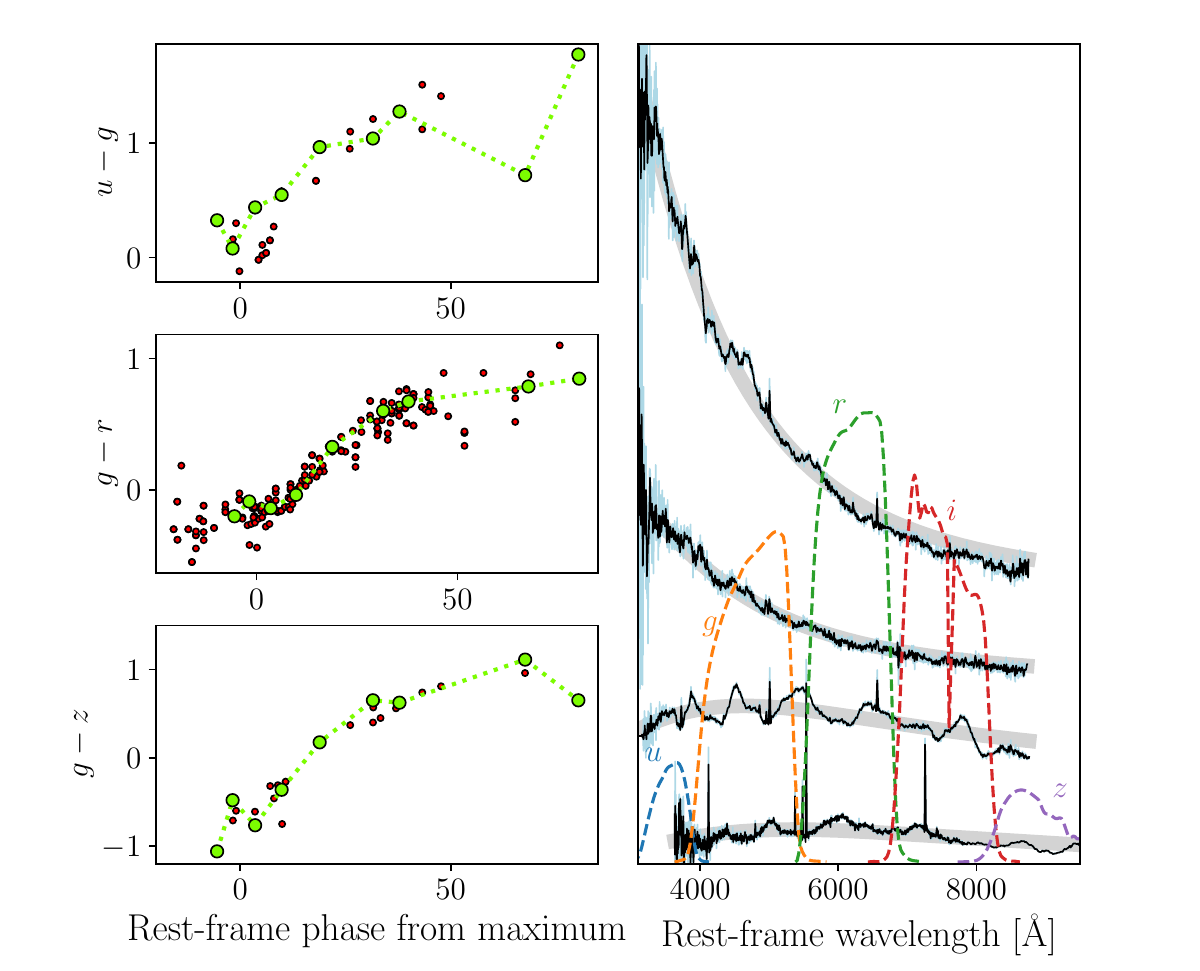}
    \caption{Left panels: rest-frame color evolution of SN~2019neq (red dots are photometric points and green dots are obtained from synthetic photometry on the observed spectra). Right panel: four representative spectra of SN~2019neq (see Fig. \ref{fig:19neq_spec}) within $\sim80$ days from the maximum luminosity. The dashed curves represent the throughput of the $u$, $g$, $r$, $i$ and $z$-passband filters.}
    \label{fig:colors}
\end{figure*}

\begin{figure}
    \centering
    \includegraphics[width=0.5\textwidth]{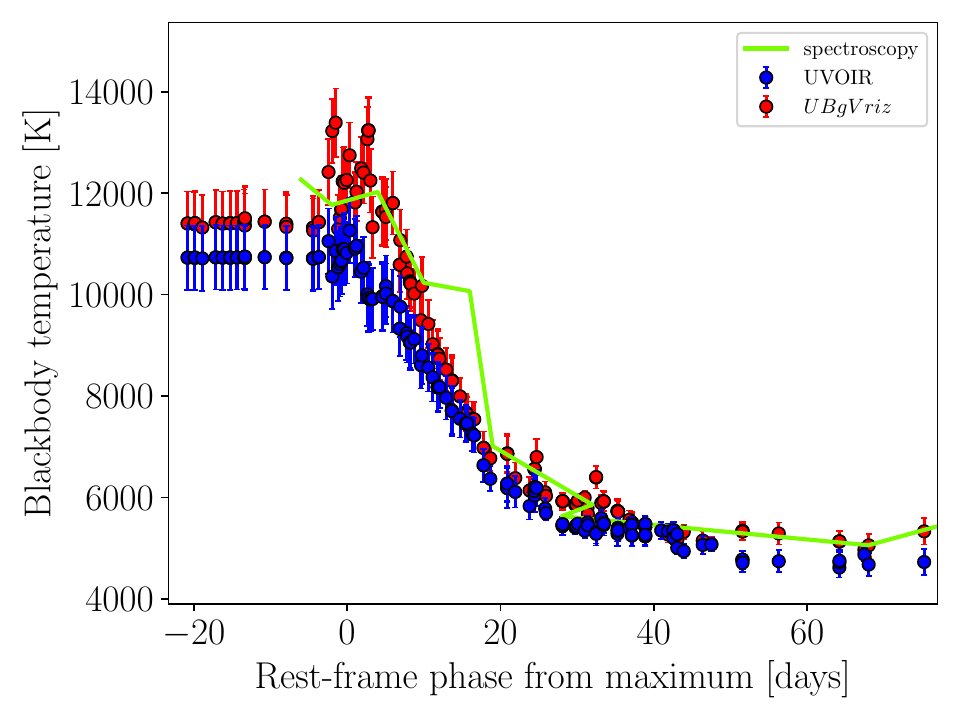}
    \caption{Evolution of the photospheric temperature. Blue dots represent values obtained integrating over the entire set of multiband photometry available, while red dots were obtained using only the $UBgVriz$ bands. The green line represents the evolution of the photospheric temperature obtained from a blackbody fit to the spectra.}
    \label{fig:bbtemp}
\end{figure}

\begin{figure}
    \centering
    \includegraphics[width=0.5\textwidth]{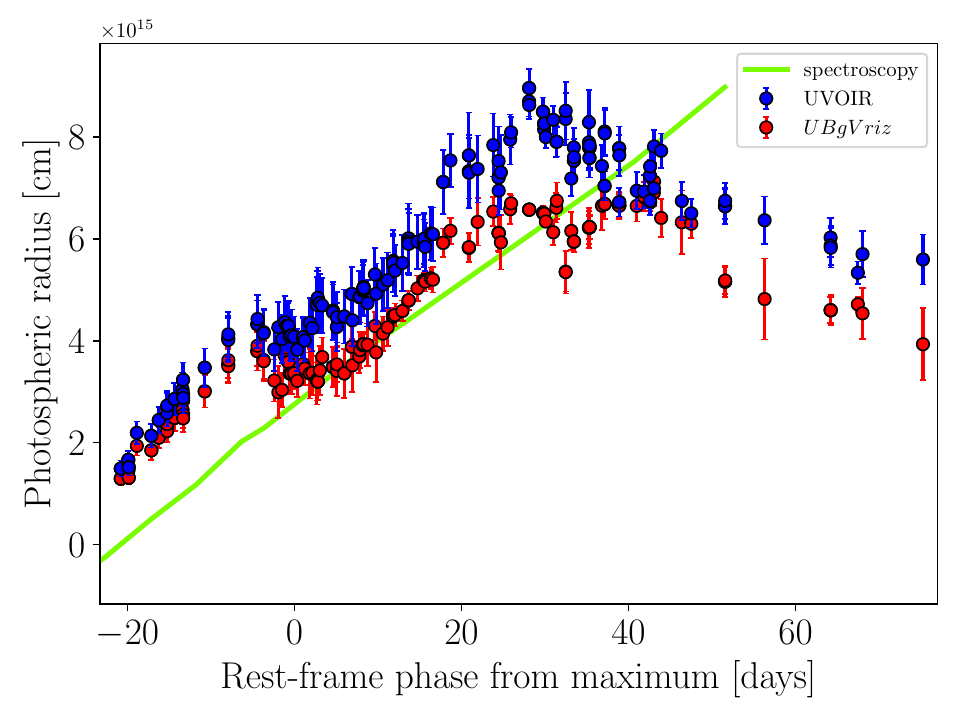}
    \caption{Evolution of the photospheric radius. Dots are color-coded as in Fig.~\ref{fig:bbtemp}. The green line was obtained using the photospheric velocities measured with the spectra.}
    \label{fig:bbrad}
\end{figure}

We discuss the SED evolution by analyzing the time variation of properties deduced from the multicolor photometry presented in Sec.~\ref{sec:photometry}. In detail, we present the rest-frame colours (Fig.~\ref{fig:gmr},~\ref{fig:colors}), photospheric temperature and radius evolution. Photospheric temperatures were estimated by fitting a blackbody curve to the SED and the photospheric radius was inferred via the Stefan-Boltzmann law. To exclude any possible UV/NIR deviations from the blackbody, we repeated the calculation excluding the $uvw2,uvm2,uvw1,J,H,K_{\rm s}$-fluxes from the SED (Figs.~\ref{fig:bbtemp}, \ref{fig:bbrad}). In addition, for each epoch $t=\phi+24$ (hence the phase from explosion, assuming an explosion phase of $-24$ days, see Sec.~\ref{sec:obslcs}) we estimated the photospheric temperature and radius from spectra: to do this, we fitted blackbody curve to spectra and computed for each epoch $t$ the radius as $v_{\rm phot}(t) \times t$ \citep[see][and discussion in Sec.~\ref{sec:19neq_phvel}]{dessartetal2015}. However, we found no major temperature differences between the three estimates and conclude that, at least at photospheric post-maximum phases (see later), the SED is reasonably well described by a blackbody over UV/optical/NIR wavelengths.

We compare the color evolution of SN~2019neq with other SLSNe~I with different evolutionary timescales. After the initial nearly-constant phase, the colours of SN~2019neq rapidly evolve towards red similar to the fastest-evolving SLSNe~I of this sample \citep[see e.g. PTF11rks, SN 2010gx and SN 2020ank,][]{inserraetal2013,pastorelloetal2010,kumaretal2020}. In contrast, colours of slower evolving SLSNe I redden at a slower pace compared to the fastest ones; the behaviour of SN 2017gci is odd, but its atypical color evolution is likely ascribable to the $\sim0.6$-mag-wide bumps in its LC \citep{fioreetal2021}. However, in all cases SLSNe I display blue colours around the pre-maximum/early-post-maximum phases \citep{pastorelloetal2010,quimbyetal2011,chomiuketal2011,leloudasetal2012,inserraetal2013}. In particular, at this phase the rest-frame $g-r$ colour remains almost constant at $\sim{-0.25}$ mag for every SLSNe I considered in Fig. \ref{fig:gmr}; which likely means that the SED does not significantly evolve. This is true also for SN~2019neq: to verify whether this behaviour is seen in other color indexes, we extend the time coverage of the observed photometry by performing synthetic photometry on the spectra (see dotted lines in Fig.~\ref{fig:colors}), this is particularly useful in $u$ band (although the flux calibration of the spectra is less precise and synthetic-photometry measurements are in part extrapolated, see Fig.~\ref{fig:colors}, right panel). As can be seen in Fig.~\ref{fig:colors}, the initial flattening appears present in $u-g$ and $g-z$ too, although very early photometric measurements can be done only in $g$ and $r$ filters. Subsequently, we see that after $\sim30-40$ days from the maximum luminosity the $g-r$ colour curve flattens again. 

The nearly constant initial behaviour of the color could reflect the early photospheric-temperature evolution; however, due to the lack of UV data at pre-maximum phases, the initial plateau phase cannot be fully trusted. Soon after maximum, the photospheric temperature starts to decline, reaching a new constant value after $\sim30$ days \citep[similar to other SLSNe I, see also][]{nicholletal2017}. Interestingly, at comparable phases, the photospheric radius reaches a maximum (Fig.~\ref{fig:bbrad}) and then starts to recede: a similar trend of the photospheric-radius evolution is described by a simple model \citep{liuetal2018} assuming homologous expansion and constant opacity. However, more detailed calculations of the photospheric radius with a proper time-varying opacity are needed to constrain the ejecta-density profile of SN~2019neq, and this will be the subject of future investigations. 
\begin{table}
    \centering
    \begin{tabular}{llll}
\hline
Name&$g_{\mathrm{norm},-15}$ &$g_\mathrm{max}$&Reference\\
\hline
SN2018bym&0.26(0.05)&$-21.88$(0.02)&\citet{chenetal2023a}\\
SN2018fcg&1.55(0.05)&$-20.32$(0.05)&\citet{chenetal2023a}\\
SN2018kyt&1.73(0.00)&$-20.68$(0.07)&\citet{chenetal2023a}\\
SN2019cdt&0.74(0.01)&$-21.07$(0.07)&\citet{chenetal2023a}\\
SN2019aamp&0.20(0.04)&$-22.14$(0.02)&\citet{chenetal2023a}\\
SN2019kwt&0.38(0.07)&$-22.57$(0.02)&\citet{chenetal2023a}\\
SN2019eot&0.17(0.07)&$-22.43$(0.03)&\citet{chenetal2023a}\\
SN2019lsq&0.54(0.05)&$-20.68$(0.03)&\citet{chenetal2023a}\\
SN2019nhs&0.36(0.05)&$-21.48$(0.02)&\citet{chenetal2023a}\\
SN2019stc&0.43(0.01)&$-20.39$(0.03)&\citet{chenetal2023a}\\
SN2019unb&0.18(0.08)&$-19.91$(0.08)&\citet{chenetal2023a}\\
SN2019ujb&0.25(0.10)&$-21.42$(0.08)&\citet{chenetal2023a}\\
SN2019zbv&0.36(0.07)&$-21.99$(0.03)&\citet{chenetal2023a}\\
SN2020fvm&0.29(0.05)&$-21.43$(0.06)&\citet{chenetal2023a}\\
SN2020auv&0.35(0.06)&$-21.72$(0.05)&\citet{chenetal2023a}\\
SN2020exj&0.91(0.03)&$-20.48$(0.02)&\citet{chenetal2023a}\\
SN2020htd&0.34(0.10)&$-21.39$(0.02)&\citet{chenetal2023a}\\
SN2019neq&1.10(0.13)&$-21.62$(0.20)&this work\\
SN2010gx&1.16(0.02)&$-20.91$(0.20)&\citet{pastorelloetal2010}\\
PTF11rks&4.41(0.02)&$-21.00$(0.05)&\citet{inserraetal2013}\\
SN2015bn&0.07(0.05)&$-22.17$(0.04)&\citet{nicholletal2016}\\
PTF12dam&0.08(0.07)&$-21.79$(0.10)&\citet{nicholletal2013}\\
SN2011ke&0.28(0.03)&$-21.60$(0.05)&\citet{inserraetal2013}\\
SN2020wnt&0.09(0.03)&$-20.44$(0.03)&\citet{gutierrezetal2022}\\
\hline
    \end{tabular}
    \caption{The SLSNe I sample used in this work (including SN 2019neq) with the values of $g_{\mathrm{norm},-15}$ and $g_\mathrm{max}$ and the reference used for the data (errors in parenthesis). Absolute magnitudes in AB system, corrected for foreground extinction and K-corrected, and referred to the cosmological model assumed in the present work.}
    \label{tab:ztfsample}
\end{table}

\subsection{Photospheric velocity}
\label{sec:19neq_phvel}
\begin{figure}
    \centering
    \includegraphics[width=0.5\textwidth]{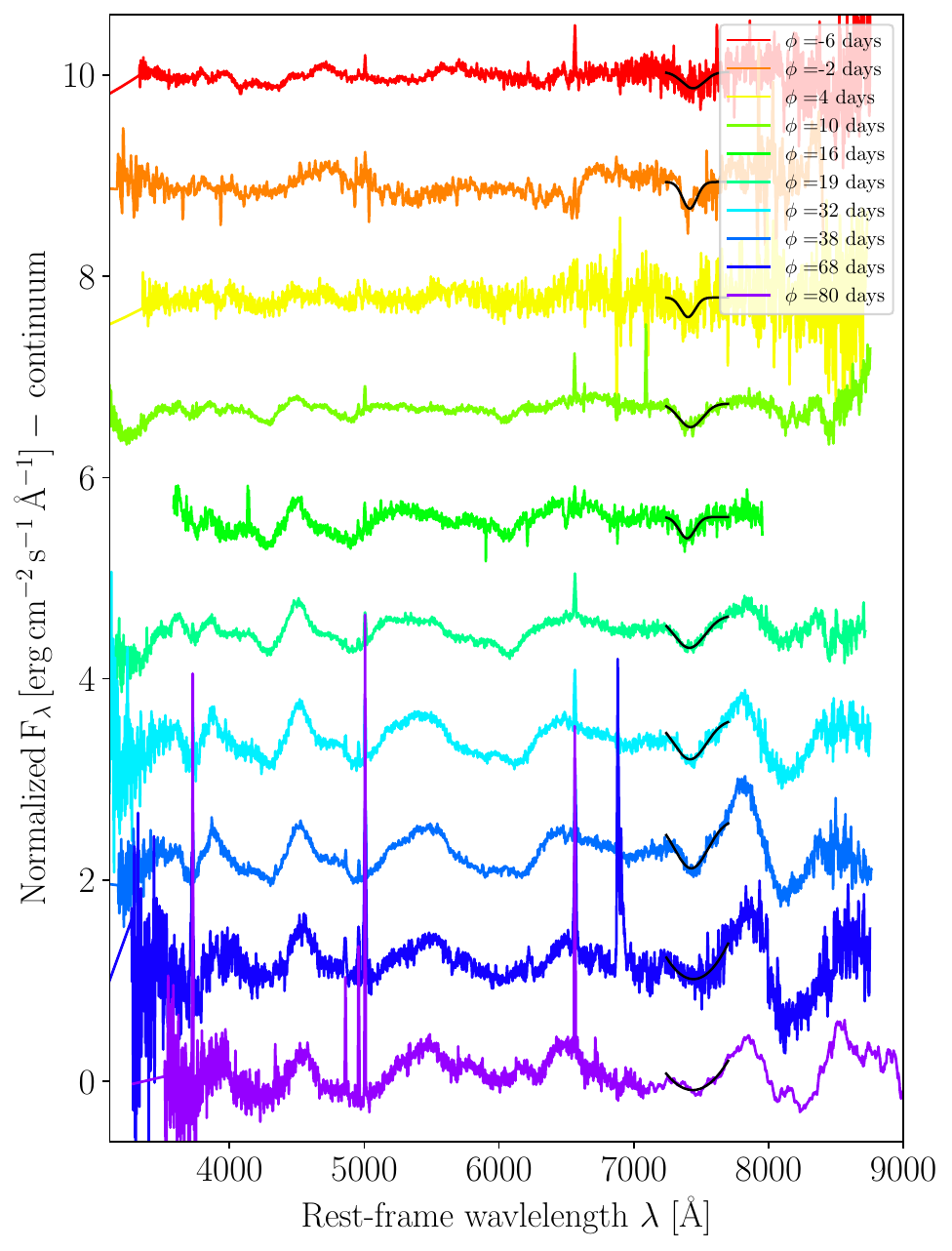}
    \caption{Normalized and continuum subtracted spectra of SN~2019neq, with the photospheric velocity being measured by fitting the minimum of the \ion{O}{I} $\lambda7774$ absorption feature with a gaussian (black solid lines).}
    \label{fig:vphot1}
\end{figure}

\begin{figure}
    \centering
    \includegraphics[width=0.5\textwidth]{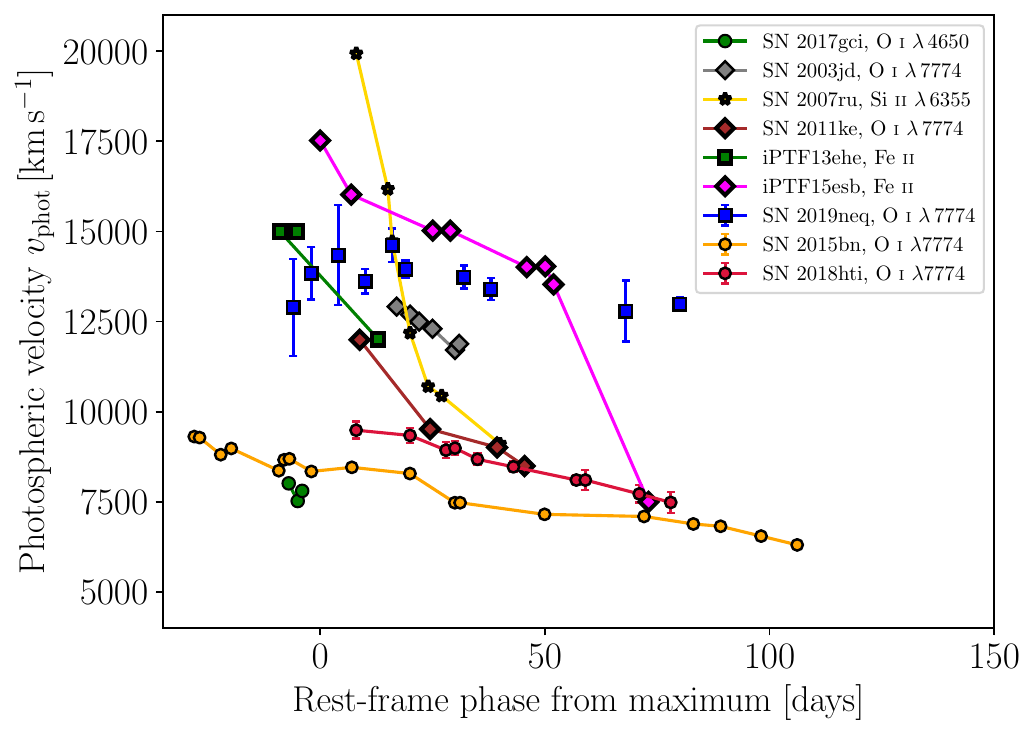}
    \caption{Photospheric velocities of SN~2019neq (blue squares) compared with those of SN~2003jd \citep[silver diamonds,][]{valentietal2008}, SN 2007ru \citep[yellow stars,][]{sahuetal2009}, SN 2011ke \citep[purple diamonds,][]{inserraetal2013}, iPTF13ehe \citep[green squares,][]{yanetal2015}, iPTF15esb \citep[magenta diamonds,][]{yanetal2017}, SN~2015bn \citep[orange dots,][]{nicholletal2016}, SN~2017gci \citep[green dots,][]{fioreetal2021} and SN~2018hti \citep[red dots,][]{fioreetal2022}}.
    \label{fig:vphot2}
\end{figure}

To estimate the photospheric velocity of SN~2019neq, the minima of the \ion{O}{I} $\lambda\,7774$ P-Cygni profiles were measured with a gaussian fit (see Fig.~\ref{fig:vphot1}) after having normalized and continuum-subtracted the spectra. These measurements were performed on the spectra at phases $-6$ to 80 days with respect to the maximum luminosity. By measuring the Doppler shift compared to the rest-frame wavelength of the corresponding maximum, we obtained the expansion velocity. This was used as a proxy for the photospheric velocity \citep{dessartetal2015} (see Fig. \ref{fig:vphot2}). Errorbars were estimated by changing the continuum level multiple times before performing the fit. The comparison between the radii deduced from photometry and spectroscopy confirms that the \ion{O}{i} is a good tracer of the photosphere for SN~2019neq for $\phi$ between $\sim-5$ and $18$ days, where we inferred an average photospheric velocity of $\sim12500\,\mathrm{km\,s^{-1}}$. Overall, the Doppler shift measured with respect to the rest-frame wavelength of the emissions corresponds to a photospheric velocity $v(\mathrm{O\,{\scriptsize I}})\approx12500-15000\,\mathrm{km\,s^{-1}}$. We also compared the photospheric velocities of SN~2019neq with those of other SLSNe~I whose velocity measurements are available in literature: iPTF13ehe \citep{yanetal2015}, iPTF15esb \citep{yanetal2017}, SN~2015bn \citep{nicholletal2016}, SN 2017gci \citep{fioreetal2021} and SN 2018hti \citep{linetal2020,fioreetal2022}. In addition, given the unusually high velocities of SN~2019neq, we included in the comparison sample also two SNe Ic BL, namely SN 2003jd \citep{valentietal2008} and SN 2007ru \citep{sahuetal2009} (see Fig.~\ref{fig:vphot2}). The high velocities of SN~2019neq do not seem to be correlated with a steeper average velocity gradient compared to slower SLSNe I (like SN~2015bn or SN~2018hti), as suggested by \citet{inserraetal2018b},\citep{koenivestothandvinko2021}, whereas this remains true for iPTF13ehe and iPTF15esb. The same appears to be also valid for the SNe Ic BL SN 2007ru and SN 2003jd, the latter reaching even higher velocities at maximum epochs. We notice that the velocities of SN~2019neq deduced by us are in tension with those deduced by \citet{koenivestothandvinko2021}, who measure $v_{\rm phot}\simeq23000\,\mathrm{km\,s^{-1}}$ at $\phi=-4$ days. However, \citet{koenivestothandvinko2021} used a different method to infer the $v_{\rm phot}$ values: they cross-correlated the observed spectra of SN~2019neq with a template \textsc{syn++} spectrum computed with $v_{\rm phot}=10000\,\mathrm{km\,s^{-1}}$ and found a velocity difference $\Delta v_{\rm X}$. The $\Delta v_{\rm X}$ values were then used to obtain the $v_{\rm phot}$ measurements after having applied a correction \citep[see Section 4.1 and equations 10, 11 in][]{koenivestothandvinko2021}.
\subsection{Modelling SN~2019neq}
\label{sec:19neq_model}
We used the nebular spectrum of SN~2019neq at 231 days to infer an estimate of the mass of the oxygen zone of the ejecta. The absolute integrated luminosity emitted within the \ion{O}{i} $\lambda\,7774$ feature can be used to infer the O-zone mass \citep[see equations 7,8 of][]{jerkstrandetal2017}. In addition, we used the \textsc{sumo} models for the nebular spectra\footnote{\texttt{https://zenodo.org/records/5578729}.} of SNe Ic \citep{jerkstrandetal2017}. The broad constraint obtained on the ejecta mass will be used in Sec.~\ref{sec:mosfit} as prior to favour one of the possible best-fit bolometric LCs obtained with the \textsc{mosfit} \citep{guillochonetal2017,guillochonetal2017soft} tool. For this same SN, \citet{koenivestothetal2020} estimated an ejecta mass of $\sim23\,\mathrm{M_\odot}$ assuming an ejecta opacity $\kappa=0.1\,\mathrm{cm^2\,g^{-1}}$. In the following, we will more carefully investigate this comparing the photometric best-fit value with the analysis of nebular spectroscopy.
\subsubsection{Nebular-spectrum modelling}
\label{sec:19neq_sumo}
The \textsc{sumo} nebular spectra are single-zone models computed for three different compositions (pure-O, C-burning ashes and OMg). For each composition, spectra calculations assume a phase of 400 days after the explosion, the homologous expansion of the ejecta at a constant velocity $V=8000\,\mathrm{km\,s^{-1}}$ and $N = 100$ random clumps for different ejecta masses\footnote{In the following, we will refer to the ejecta mass in unit of solar masses with which \textsc{sumo} models are computed as $M^\textsc{sumo}_\mathrm{ejecta}$ to distinguish them from other ejecta-mass estimates.} $M^\textsc{sumo}_\mathrm{ejecta}=3,10,30$, energy deposition $E_\mathrm{dep}=2.5\times10^{41}-2\times10^{42}\,\mathrm{erg\,s^{-1}}$ and filling factors $f=0.1,0.01,0.001$. The filling factor expresses the percentage volume of clumps, hence lower $f$ values correspond to more clumped ejecta. 

Before measuring the \ion{O}{i} $\lambda\,7774$ absolute luminosity, we estimated the residual contribution of the host-galaxy emission using the host-galaxy spectrum adjacent to that of SN~2019neq. Then this was scaled and subtracted from the spectrum of SN~2019neq until the narrow galaxy emission was removed from the SN spectrum. Then, we measured an integrated luminosity of the \ion{O}{I} $\lambda$ 7774 line of $L_{7774}\lesssim2\times10^{40}\,\mathrm{erg\,s^{-1}}$. {Due to the relatively low signal-to-noise ratio between 7000-8000 \AA{}, we assume this value as an upper limit}. Similar to the \textsc{sumo} models, we assumed a maximum velocity of the fluid elements $V=8000\,\mathrm{km\,s^{-1}}$. Aside this parameter, the only tunable quantities are the clumping factor $f$ and the electron fraction $x_e$. We computed the O-zone mass for the $f$ values with which \textsc{sumo} models are computed, i.e. $f=0.1,0.01,0.001$, a phase from the explosion of $231+24\,\mathrm{days}=255\,\mathrm{days}$ (see Sec.~\ref{sec:obslcs}) and we consider a range of $x_e=0.05-0.1$ \citep[this range encompasses the typical values used in \textsc{sumo} models,][]{jerkstrandetal2017}. Results are shown in Tab.~\ref{tab:ozmass}.
\begin{table}
    \centering
    \begin{tabular}{|llll|}
        \hline
         $f$&0.1&0.01&0.001\\
         \hline
         $n_e\,[10^8\,\mathrm{cm^{-3}}]$&1.3&4.1&13\\
         $M(O)\,[\mathrm{M_\odot}]$&122(61)&39(19)&12(6)\\
         \hline
    \end{tabular}
    \caption{Electron density $n_e$ and O-zone mass $M(O)$ computed for $f=0.1,0.01,0.001$ after having fixed $L_{7774}=2\times10^{40}\,\mathrm{erg\,s^{-1}}$, $t=255$ days (here we have assumed a rise time of 24 days, see Sec.~\ref{sec:obslcs}) and $V=8000\,\mathrm{km\,s^{-1}}$. The $M(O)$ values in parentheses are computed for $x_e=0.1$, otherwise for $x_e=0.05$. }
    \label{tab:ozmass}
\end{table}
The case $f=0.1$ leads to very high O-zone masses ($M(O)>100\,\mathrm{M_\odot}$): based on the rapid LC-evolution timescales, in the following we will not consider this case anymore. The $f=0.01,0.001$ cases allow for lower O-zone masses permitted by lower-mass ejecta. In addition, our mass estimates are strongly affected by our assumption on the electron fraction $x_e$, the clumping factor $f$ and the maximum velocity of the clumps $V$.

We then consider \textsc{sumo} models computed for $f=0.01,0.001$, $M^\textsc{sumo}_{\rm ejecta}=3,10,30$ for pure-Oxygen, OMg and C-burning abundances, and different values of the energy deposition $E_{\rm dep}=2\times10^{41}-2\times10^{42}\,\mathrm{erg\,s^{-1}}$ (see Figs.~\ref{fig:sumo1}, \ref{fig:sumo2}, \ref{fig:sumo3}, \ref{fig:sumo4}, \ref{fig:sumo5}, \ref{fig:sumo6}). [\ion{O}{i}] seems to be better described by models with $M_{\rm ejecta}^\textsc{sumo}$=10 and $f=0.01$ for higher energy deposition, thus possibly favouring higher values of the electron fraction $x_e$. Other models underpredict the [\ion{O}{i}] $\lambda$ 6300,6364 feature, except the C-burning models  with $M_{\rm ejecta}^\textsc{sumo}=30$ and $E_{\rm dep}>1\times10^{42}\,\mathrm{erg\,s^{-1}}$. Furthermore, OMg-composition models seem less suitable to explain the \ion{O}{i} features seen in the nebular spectrum of SN~2019neq, in particular for $E_{\rm dep}=2\times10^{41}\,\mathrm{erg\,s^{-1}}$ (see Figs.~\ref{fig:sumo5},\ref{fig:sumo6}). However, the temporal discrepancy between the observed nebular spectrum of SN~2019neq and the \textsc{sumo} solutions makes the \textsc{sumo} spectra by a factor $(400/255)^3\sim3.9$ less dense than the observed, if ejecta mass and expansion velocity are the same: this makes their spectral comparisons less relevant to discriminate among the different ejecta configurations. In addition, we warn the reader that a more careful estimate of the physical parameters of SN 2019neq would need independent constraints to break some degeneracies (e.~g. between $n_e$ and $f$ or between $E_{\rm in}$ and $M_{\rm ejecta}$); this is however beyond the scope of the present work.
\subsubsection{Light-curve modelling}
\label{sec:mosfit}
We modelled the LCs of SN~2019neq to estimate the physical parameters of the explosion assuming the magnetar and the ejecta-CSM interaction. We used the Modular Open Source Fitter for Transients (\textsc{mosfit}) tool \citep{guillochonetal2017,guillochonetal2017soft} which offers a set of modules to fit the observed multicolour LCs for different kind of transients. For SN~2019neq, we ran \textsc{mosfit} using the \textsc{slsn} and the \textsc{csm} modules, computing synthetic LCs powered by magnetar spindown \citep{kasenandbildsten2010,inserraetal2013} and CSM \citep{chatzopoulosetal2012} interaction, respectively. We fixed $\kappa=0.1\,\mathrm{cm^2\,g^{-1}}$ for both models which is suitable for SNe Ic \citep[e.g.][]{nagy2018}.

\begin{figure}
    \centering
    \includegraphics[width=0.4\textwidth]{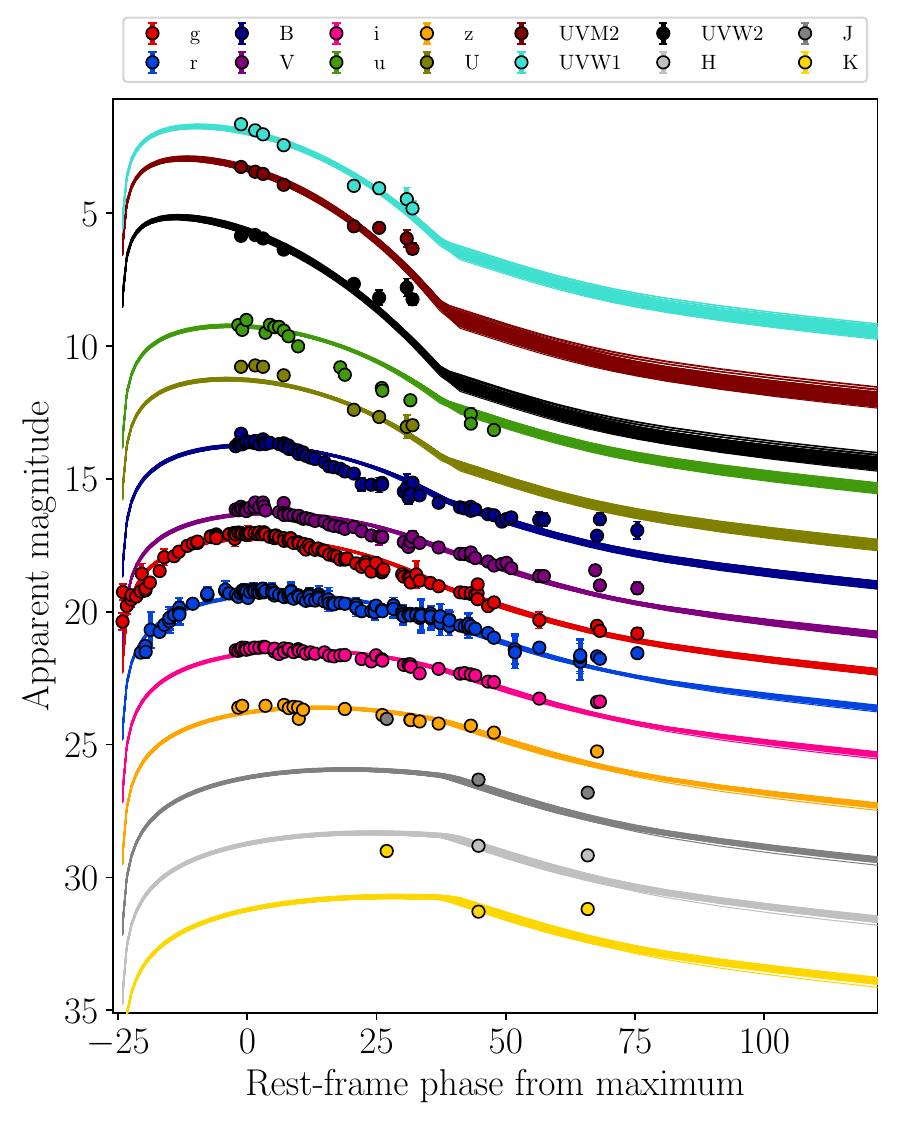}
    \caption{Multicolor synthetic LCs output by \textsc{mosfit} \textsc{slsn}-model.}
    \label{fig:slsn_mosfit}
\end{figure}

\begin{figure}
    \centering
    \includegraphics[width=0.4\textwidth]{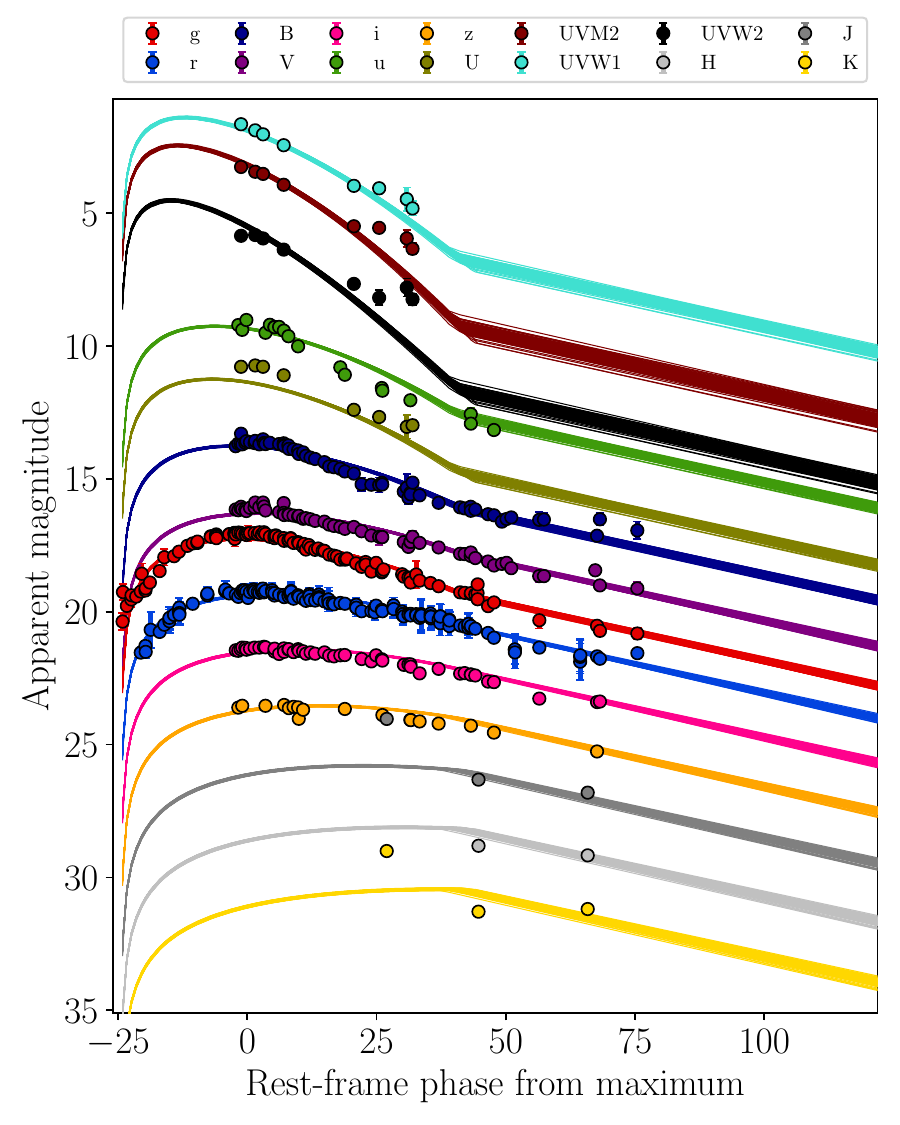}
    \caption{Same as Fig.~\ref{fig:csm_mosfit}, but for the \textsc{csm}-model.}
    \label{fig:csm_mosfit}
\end{figure}

The \textsc{csm} fit was set up adopting a broken power-law ejecta density profile as in \citet{chatzopoulosetal2012} with fixed exponents $n=12$ and $s=2$, suitable for a steady wind-like CSM. A shell-like \textsc{csm} ($s=0$) fit is not well constrained by the data, hence we excluded it. The $s=0$ \textsc{csm} best-fit parameters are a CSM mass of $M_{\rm CSM}\simeq0.7\,\mathrm{M_\odot}$, a progenitor radius $R_0\simeq1.4\times10^{14}\,\mathrm{cm}$, a CSM density $\rho\simeq1.8\times10^{-10}\,\mathrm{g\,cm^{-3}}$, an ejecta mass $M_{\rm ejecta}\simeq21.9\,\mathrm{M_\odot}$, an average ejecta velocity $v_{\rm ejecta}\simeq4100\,\mathrm{km\,s^{-1}}$ and a temperature floor $T_{\rm min}\simeq3700\,\mathrm{K}$. This corresponds to a kinetic energy of $\sim3.7\times10^{51}\,\,\mathrm{erg}$. The ejecta mass for the \textsc{csm} fit broadly agrees with the \textsc{sumo} models computed $M^\textsc{sumo}_\mathrm{ejecta}=30$. The \textsc{slsn} fit predicts a lower ejecta mass ($\sim12\,\mathrm{M_\odot}$), thus in better agreement with \textsc{sumo} spectra with $M^\textsc{sumo}_\mathrm{ejecta}=10-30$. This is especially true for \textsc{sumo} models computed with $M_{\rm ejecta}^\textsc{sumo}=10$ and $f=0.01$. To verify whether higher ejecta-mass models return reasonable results, we performed another \textsc{slsn} fit, this time fixing an ejecta mass $M_\mathrm{ejecta}=25\,\mathrm{M_\odot}$ and fitting the opacity: not surprisingly, in this last case the best-fit opacity is lower ($\kappa=0.05\,\mathrm{cm^2\,g^{-1}}$), but however reasonable in the case of H/He-poor events \citep{nagy2018}. Both the $\kappa=0.05\,\mathrm{cm^2\,g^{-1}}$ and the $\kappa=0.1\,\mathrm{cm^2\,g^{-1}}$ \textsc{slsn} fits provide ejecta-mass estimates that are broadly consistent with what can be inferred via the nebular-spectrum interpretation presented in the previous Section. The other best-fit parameters for the two \textsc{slsn} fits are a magnetic field $B\simeq6\times10^{14}\,\mathrm{G}$, an initial spin period $P_{\rm init}\sim1.1-1.5\,\mathrm{ms}$, a temperature floor $T_{\rm min}\simeq6000\,\mathrm{K}$ and an average ejecta velocity $v_{\rm ejecta}\simeq10000\,\mathrm{km\,s^{-1}}$. This correspond to a kinetic energy of $1.3-2.7\times10^{52}\,\mathrm{erg}$. The best-fit parameters for the \textsc{csm} and \textsc{slsn} are listed in Tab.~\ref{tab:mosfitcsm},\ref{tab:mosfitslsn} and the corner plots showing their posterior distributions are shown in Fig.~\ref{fig:corner_slsn}, \ref{fig:corner_csm}, respectively.
\begin{table*}
    \centering
    \begin{tabular}{lllllll}
        \hline
         $M_{\rm CSM}\,[\mathrm{M_\odot}]$&$R_0\,[10^{14}\,\mathrm{cm}]$&$\rho\,[\mathrm{g\,cm^{-3}}]$&$M_{\rm ejecta}\,[\mathrm{M_\odot}]$&$\kappa\,[\mathrm{cm^2\,g^{-1}}]$&$T_{\rm min}\,[10^3\,\mathrm{K}]$&$v_{\rm ejecta}\,[10^3\,\mathrm{km\,s^{-1}}]$ \\
         \hline
         0.74&1.36&1.78e-10&21.88&$0.1(^{*})$&3.73&4.05\\
         \hline
    \end{tabular}
    \caption{\textsc{mosfit csm} physical parameters (the opacity, marked with $(^{*})$ is a fixed parameter; the other ones are best-fit values).}
    \label{tab:mosfitcsm}
\end{table*}
\begin{table*}
    \centering
    \begin{tabular}{lllllll}
        \hline
         $B\,[10^{14}\,\mathrm{G}]$&$M_{\rm NS}\,[\mathrm{M_\odot}]$&$P_{\rm spin}\,[\mathrm{ms}]$&$M_{\rm ejecta}\,[\mathrm{M_\odot}]$&$\kappa\,[\mathrm{cm^2\,g^{-1}}]$&$T_{\rm min}\,[10^3\,\mathrm{K}]$&$v_{\rm ejecta}\,[10^3\,\mathrm{km\,s^{-1}}]$ \\
         \hline
         5.98&1.66&1.47&11.75&$0.1(^{*})$&5.94&10.05\\
         5.76&1.49&1.01&$25(^{*})$&0.05&5.96&10.43\\
         \hline
    \end{tabular}
    \caption{\textsc{mosfit slsn} physical parameters (the one with $(^{*})$ is a fixed parameter, the other ones are best-fit values).}
    \label{tab:mosfitslsn}
\end{table*}
The \textsc{slsn} and \textsc{csm} LCs are shown in Fig.~\ref{fig:slsn_mosfit}, \ref{fig:csm_mosfit} and are able to capture the pre-maximum/maximum luminosity epochs. The early post-maximum photometric detections in $uvm2$ and $uvw2$ bands cannot be properly described by both models; moreover, red and NIR-bands photometric points at a phase later than $\sim+60$ days are better described by the \textsc{csm} fit. However, the UV/NIR photometric coverage is too scarce to use the previous argument as discriminating factors between the two models. We notice also that the $T_{\rm min}$ and the $v_{\rm ejecta}$ values predicted by the \textsc{slsn} model are in better agreement with the measured data (Figs.~\ref{fig:bbtemp}, \ref{fig:vphot2}) compared to those predicted in the \textsc{csm} case. 

Moreover, the spectra of SN~2019neq do not display the prominent narrow/multicomponent emission features seen e. g. in the luminous  type-IIn SN 2006gy and SN 2006tf \citep{smithetal2007,smithetal2008,kieweetal2012}. Based on these reasons, we disfavour the CSM-interaction scenario for SN~2019neq and suggest that a millisecond magnetar endowed with a magnetic field $B\simeq6\times10^{14}\,\mathrm{G}$ is the engine driving the observed luminosity of SN~2019neq. CSM interaction might have contributed during pre-/post-maximum epochs and resulted in the possible complexities of the LCs of SN 2019neq. However, we cannot rule out that more peculiar (e. g. disk-like) CSM configurations might have reprocessed the UV/X-ray photons before reaching the photosphere, thus quenching the narrow/multicomponent line formation \citep{andrewsandsmith2018}. 
\section{Conclusions}
We analyzed the spectrophotometric observations of SN~2019neq, which is among the fastest-evolving SLSNe I ever observed. Its LCs rise towards maximum in about $24$ rest-frame days and decline at a $\sim2.6$ times slower rate. Similar to other SLSNe~I, SN~2019neq displays approximately constant blue colours during the pre-maximum/maximum epochs. At these phases, the spectra display W-shaped absorptions attributed to \ion{O}{ii}, and at a phase 16-19 days remind the spectra of SNe Ic/Ic BL at maximum luminosity. From the narrow lines attributed to the host galaxy, we inferred a metallicity $Z\sim0.4\,Z_\odot$, a SFR $\sim2.6\,\mathrm{M_\odot\,yr^{-1}}$ and a sSFR of $\sim2.3\,\mathrm{Gyr^{-1}}$ at the SN site. A late nebular spectrum of SN~2019neq was scaled in absolute specific luminosity and compared with different models computed with the \textsc{sumo} single-zone, radiative-transfer code designated for nebular spectra of SNe Ic; we also used a set of analytical relations \citep{jerkstrandetal2017} to estimate the ejecta O mass. In addition, multicolor LCs of SN~2019neq were modelled with the \textsc{mosfit} tool using the \textsc{csm} and \textsc{slsn} modules to test the ejecta-CSM interaction and the magnetar scenarios, respectively. Based on the best-fit results and using the inferred nebular properties as a broad prior, we conclude that the spindown radiation of a millisecond, young magnetar with a magnetic field of $\sim6\times10^{14}$ G likely powered the luminosity of SN~2019neq at around the maximum luminosity. Given the degeneracy between the ejecta mass $M_{\rm ejecta}$ and the ejecta opacity $\kappa$, it is hard to retrieve reasonable estimates, hence we provide for them the following ranges: $M_{\rm ejecta}\simeq10-30\,\mathrm{M_\odot}$ and $\kappa\simeq0.05-0.1\,\mathrm{cm^2\,g^{-1}}$. In addition, we cannot rule out that the ejecta of SN~2019neq have interacted with a previously-ejected CSM, but either this requires a peculiar CSM topology, or that its contribution is subdominant in powering the SN around maximum luminosity.

Due to their rapid photometric evolution, SLSNe I like SN~2019neq are challenging to follow up, in particular at nebular epochs. Their careful study warrants a strong observational effort which future wide-field surveys like the Legacy Survey of Space and Time at the Vera Rubin Observatory \citep{villaretal2018} will allow for. In addition, a larger dataset is useful to improve the statistical significance of SLSNe sample studies. These are crucial to understand the diversity of SLSNe and to unravel the physical reasons for their luminous nature.
\section*{Acknowledgements}
This work is based on
observations made with the Nordic Optical Telescope, owned in collaboration by the University of Turku and Aarhus University, and operated jointly by Aarhus University, the University of Turku and the University of Oslo, representing Denmark, Finland and Norway, the University of Iceland and Stockholm University at the Observatorio del Roque de los Muchachos, La Palma, Spain, of the Instituto de Astrofisica de Canarias. The data presented here were obtained in part with ALFOSC, which is provided by the Instituto de Astrofisica de Andalucia (IAA) under a joint agreement with the University of Copenhagen and NOT. Based on observations collected at Copernico and Schmidt telescopes (Asiago, Italy) of the INAF - Osservatorio Astronomico di Padova. Based on observations made with the Gran Telescopio Canarias (GTC), installed in the Spanish Observatorio del Roque de los Muchachos of the Instituto de Astrof\'isica de Canarias, in the island of La Palma. 
A. F. acknowledges the support by the State of Hesse within the Research Cluster ELEMENTS (Project ID 500/10.006).
X. W. is supported by the National Natural Science Foundation of China (NSFC grants 12288102 \& 12033003), Scholar Program of Beijing Academy of Science and Technology (DZ:BS202002), and Tencent Xplorer Prize.
A. P. acknowledges support from PRIN-MIUR 2022.
A.M.G. acknowledges financial support by the European Union under the 2014-2020 ERDF Operational Programme and by the Department of Economic Transformation, Industry, Knowledge, and Universities of the Regional Government of Andalusia through the FEDER-UCA18-107404 grant. Y.-Z. Cai is supported by the National Natural Science Foundation of China (NSFC, Grant No. 12303054) and the International Centre of Supernovae, Yunnan Key Laboratory (No. 202302AN360001).
M. F. is supported by a Royal Society-Science Foundation Ireland University Research Fellowship. E. C., N. E. R., I. S. and A. F. acknowledge support of MIUR, PRIN 2017 (grant 20179ZF5KS) `The new frontier of the Multi-Messenger Astrophysics: follow-up of electromagnetic transient counterparts of gravitational wave sources'. S. M. acknowledges support from the Magnus Ehrnrooth Foundation and the Vilho, Yrj\"{o}, and Kalle V\"{a}is\"{a}l\"{a} Foundation. T. M. R.  acknowledges the financial support of the Vilho, Yrj{\"o} and Kalle V{\"a}is{\"a}l{\"a} Foundation of the Finnish academy of Science and Letters through the Finnish postdoc pool. I. S. is supported by fundings by the PRIN-INAF 2022 project "Shedding light on the nature of gap transients: from the observations to the models", and by the doctoral grant funded by Istituto Nazionale di Astrofisica via the University of Padova and the Italian Ministry of Education, University and Research (MIUR). N.E.R. acknowledges partial support from PRIN-INAF 2022 "Shedding light on the nature of gap transients: from the observations to the models”, from the Spanish MICINN grant PID2019-108709GB-I00 and FEDER funds, and from the program Unidad de Excelencia María de Maeztu CEX2020-001058-M. C. P. G. acknowledges financial support from the Secretary of Universities
and Research (Government of Catalonia) and by the Horizon 2020 Research
and Innovation Programme of the European Union under the Marie
Sk\l{}odowska-Curie and the Beatriu de Pin\'os 2021 BP 00168 programme,
from the Spanish Ministerio de Ciencia e Innovaci\'on (MCIN) and the
Agencia Estatal de Investigaci\'on (AEI) 10.13039/501100011033 under the
PID2020-115253GA-I00 HOSTFLOWS project, and the program Unidad de
Excelencia Mar\'ia de Maeztu CEX2020-001058.
\section*{Data availability statement}
The data presented in this paper and listed in Appendices A, B are available in the online supplementary material. The spectra will be made public via \textsc{wiserep}. 





\bibliographystyle{mnras}
\bibliography{reference} 

\appendix
\section{S-corrections and K-corrections}
\label{sec:skcorrs}
\subsection{S-correction}
\begin{figure}
    \centering
    \includegraphics[width=0.55\textwidth]{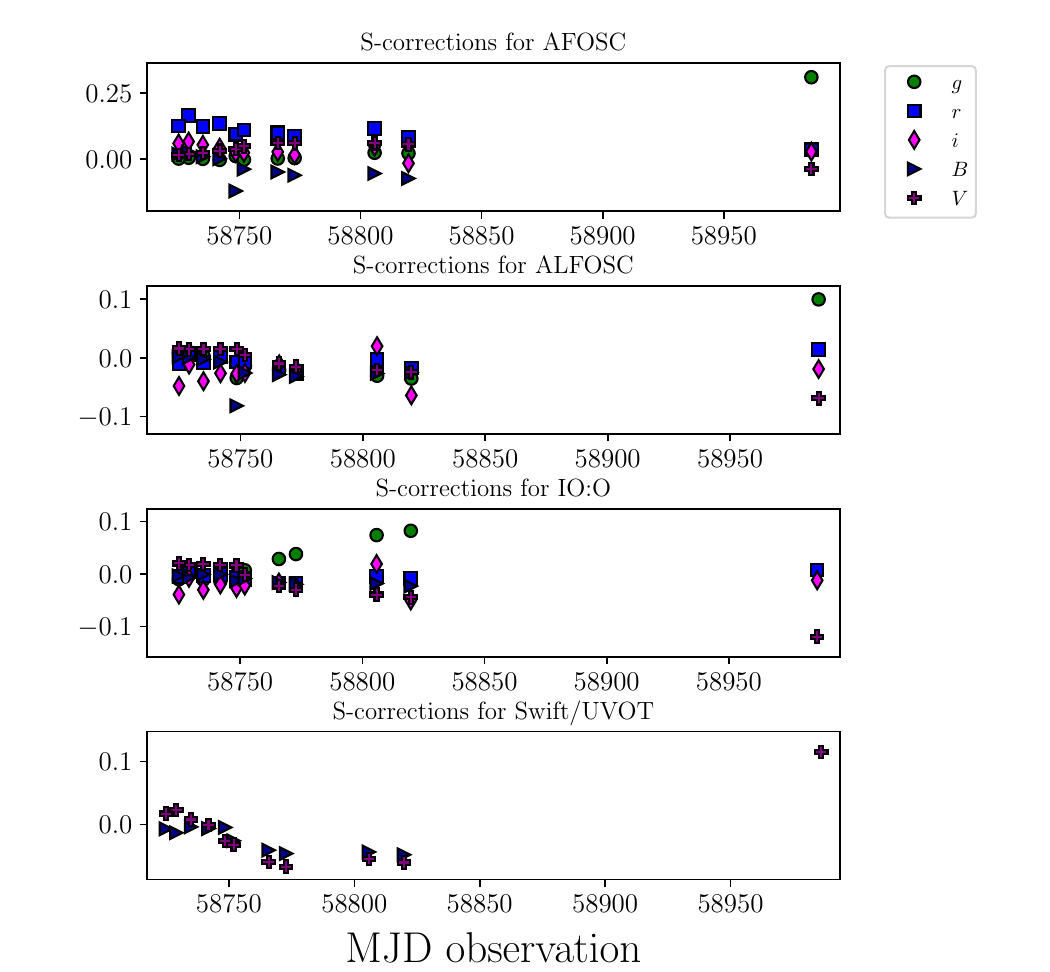}
    \caption{S-corrections for SN~2019neq for the different instrumental configurations used for the photometric follow-up of SN~2019neq.}
    \label{fig:scorrection}
\end{figure}

As different instrumental setups correspond to different photometric systems, we computed the S-correction \citep[see e.g.][]{strintzingeretal2002,pignataetal2004,eliasrosaetal2006} to bring back all the magnitude measurements to a standard system. In the present work, we define the S-correction as $S_{\rm corr}=m_{s, {\rm standard}}-m_{s,\rm instr}$, where $m_{s, {\rm standard}}$ and $m_{s,\rm instr}$ are synthetic photometry measurements performed on the spectra using the instrumental and the standard filter, respectively (see Fig.~\ref{fig:scorrection}). The S-correction values are secured in Tabs.~\ref{tab:scorrasiago},\ref{tab:scorrnot}, \ref{tab:scorrlt}, \ref{tab:scorrswift}.
These were linearly interpolated for the photometric epochs not matched by coeval spectroscopic ones. The $S_{\rm corr}$ values were then summed to the measured magnitudes and its statistical dispersion ($\lesssim0.01$ magnitudes) was propagated in our analysis (see Tab.~\ref{tab:scorrerr}). 

Finally, for the filters where the wavelength range falls at least partly outside the spectral coverage, we computed the S-correction as before, but using a blackbody as the spectrum on which synthetic photometry was performed. The maximum value for the S-correction computed using a blackbody $\Delta S_{\rm corr}$ was evaluated in two temperature ranges ($5000\,\mathrm{K}<T<8000\,\mathrm{K}$ and $8000\,\mathrm{K}<T<12000\,\mathrm{K}$, see also Sec.~\ref{sec:bbtemp}) and propagated in our analysis as an additional uncertainty due to the non-standard instrumental photometric systems.

\subsection{K-correction}
We computed K-corrections to account for the effect of the cosmological redshift on the magnitude measured in the observer frame band-pass filters. Similar to S-corrections, we computed the K-corrections performing synthetic magnitudes measurements on the spectra of SN~2019neq (see Sec.~\ref{sec:spectroscopy}) via \textsc{pysynphot}. In detail, for each band-pass filter, we derived a synthetic magnitude both for the rest-frame spectrum (for which we computed a synthetic magnitude $m_{s,{\rm rest}}$) and for the observed one (for which we computed a synthetic magnitude $m_{s,{\rm obs}}$). For each epoch and filter, the K-correction was computed as $m_{s,{\rm rest}}-m_{s,{\rm obs}}$. The resulting K-corrections are listed in Tab.~\ref{tab:19neq_kcorr}. {Similar to the $\Delta S_{\rm corr}$}, the K-corrections for the $uvw2, uvm2, uvw1, J, H, K_{\rm s}$-filter magnitudes were estimated as before but using blackbody fits to the SED.
\section{Figures}
\begin{figure*}
    \centering
    \includegraphics[width=0.9\textwidth]{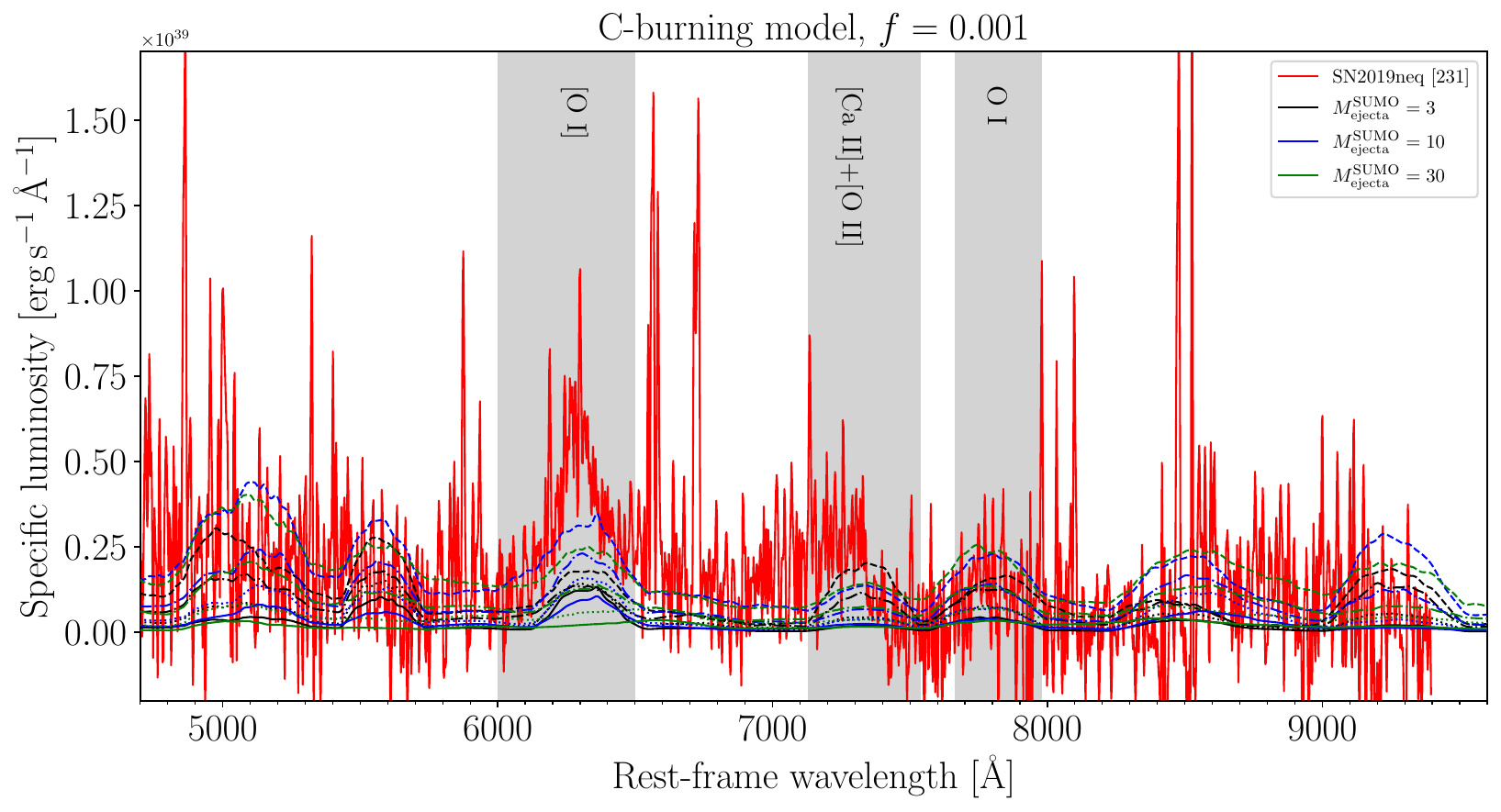}
    \caption{Comparison of the observed nebular spectrum of SN~2019neq (red line) with different \textsc{sumo} spectra computed for the C-burning model with a filling factor $f=0.001$ for different ejecta masses $M^\textsc{sumo}_\mathrm{ejecta}=10,30$ and different energy depositions $E_{\rm dep}$. Different colours correspond to different $M^\textsc{sumo}_\mathrm{ejecta}$, as labelled in the top-right corner. Solid lines correspond to an energy deposition $E_{\rm dep}=2.5\times10^{41}\,\mathrm{erg\,s^{-1}}$, dotted lines to $E_{\rm dep}=5\times10^{41}\,\mathrm{erg\,s^{-1}}$, dashed-dotted lines to $E_{\rm dep}=1\times10^{42}\,\mathrm{erg\,s^{-1}}$, dashed lines to $E_{\rm dep}=2\times10^{42}\,\mathrm{erg\,s^{-1}}$.}
    \label{fig:sumo1}
\end{figure*}
\begin{figure*}
    \centering
    \includegraphics[width=0.9\textwidth]{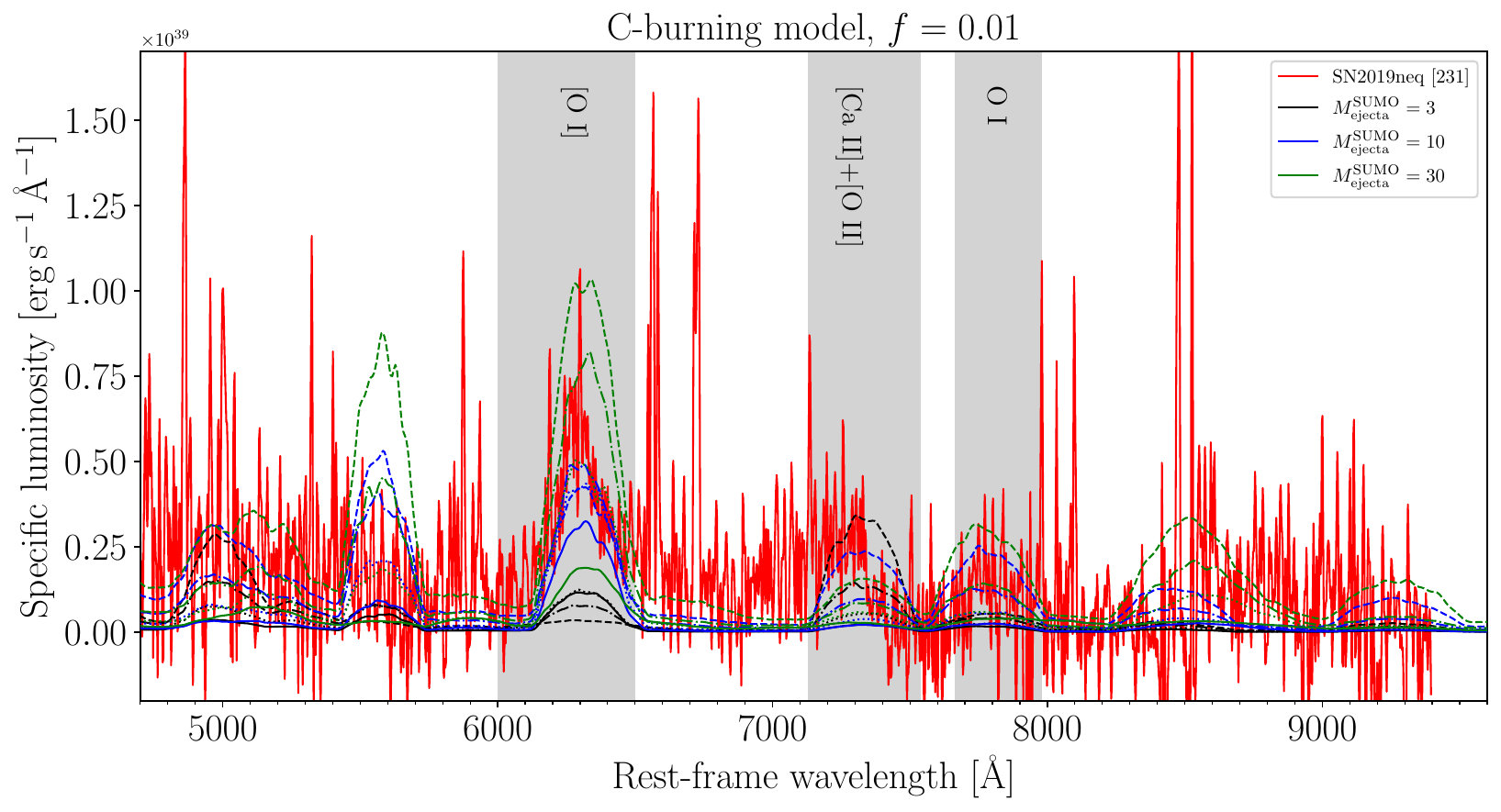}
    \caption{Similar to Fig.~\ref{fig:sumo1}, but for the C-burning model and $f=0.01$.}
    \label{fig:sumo2}
\end{figure*}
\begin{figure*}
    \centering
    \includegraphics[width=0.9\textwidth]{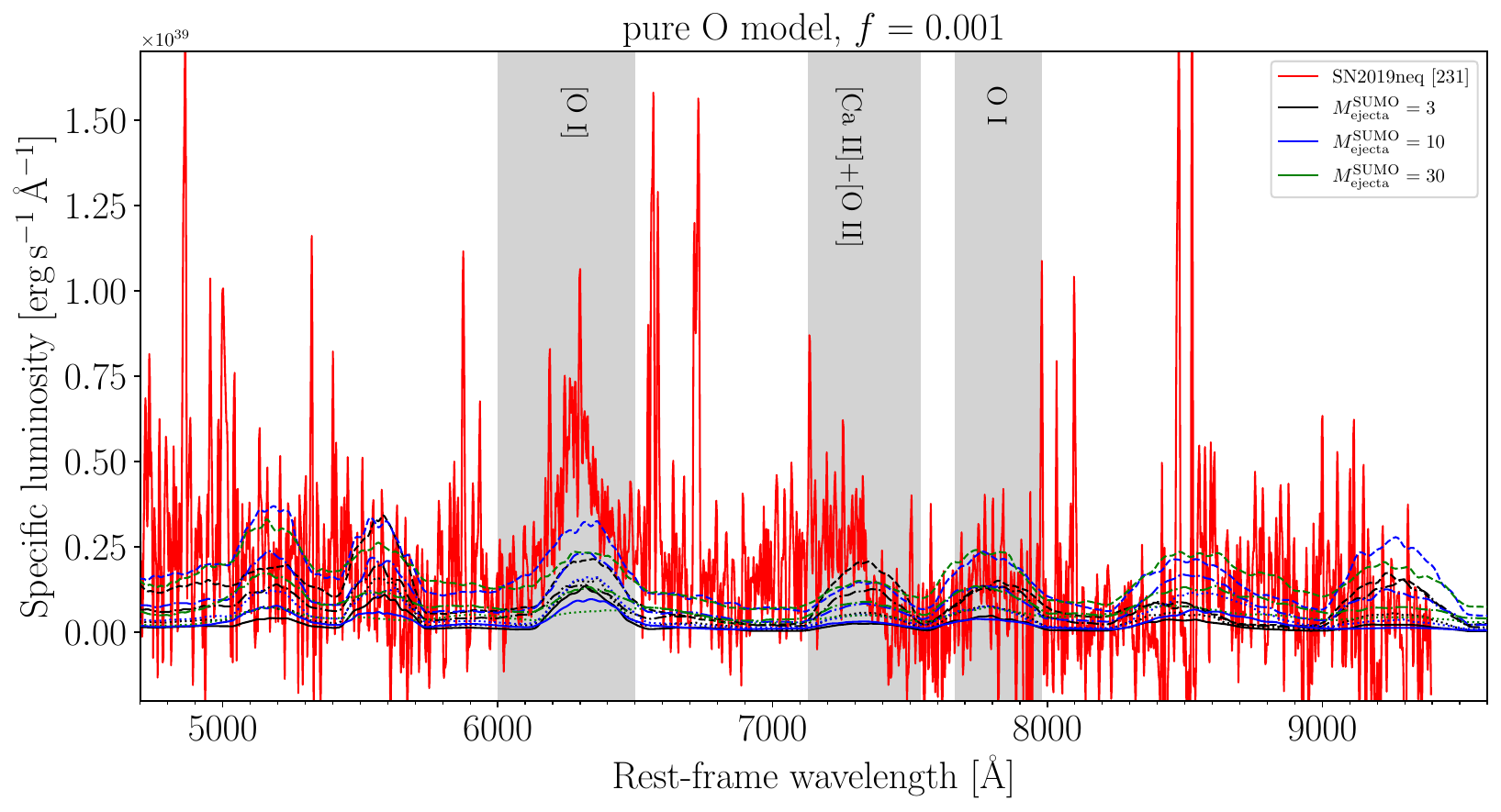}
    \caption{Similar to Fig.~\ref{fig:sumo1}, but for the pure-Oxygen model.}
    \label{fig:sumo3}
\end{figure*}
\begin{figure*}
    \centering
    \includegraphics[width=0.9\textwidth]{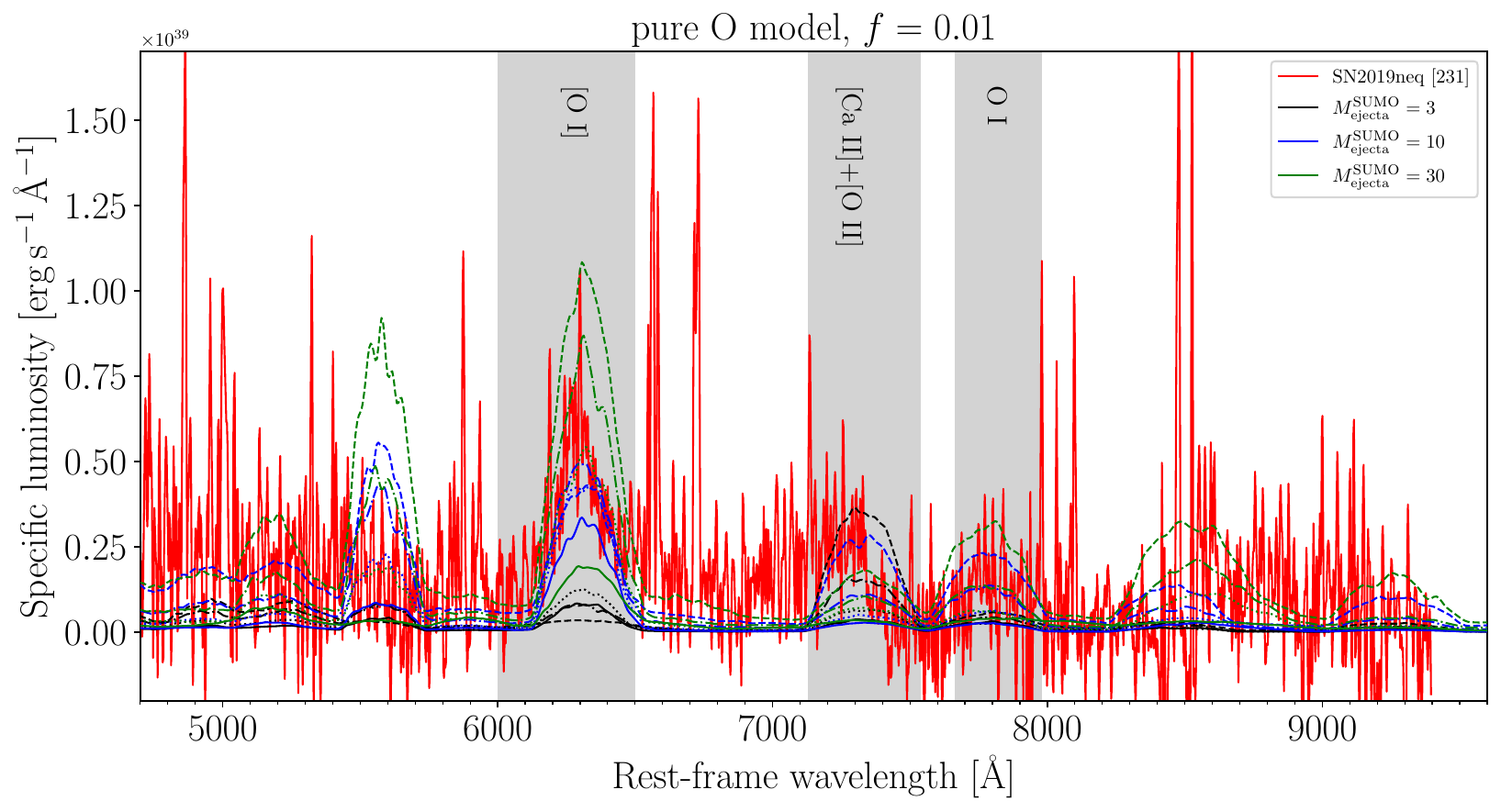}
    \caption{Similar to Fig.~\ref{fig:sumo1}, but for the pure-Oxygen model and $f=0.01$.}
    \label{fig:sumo4}
\end{figure*}
\begin{figure*}
    \centering
    \includegraphics[width=0.9\textwidth]{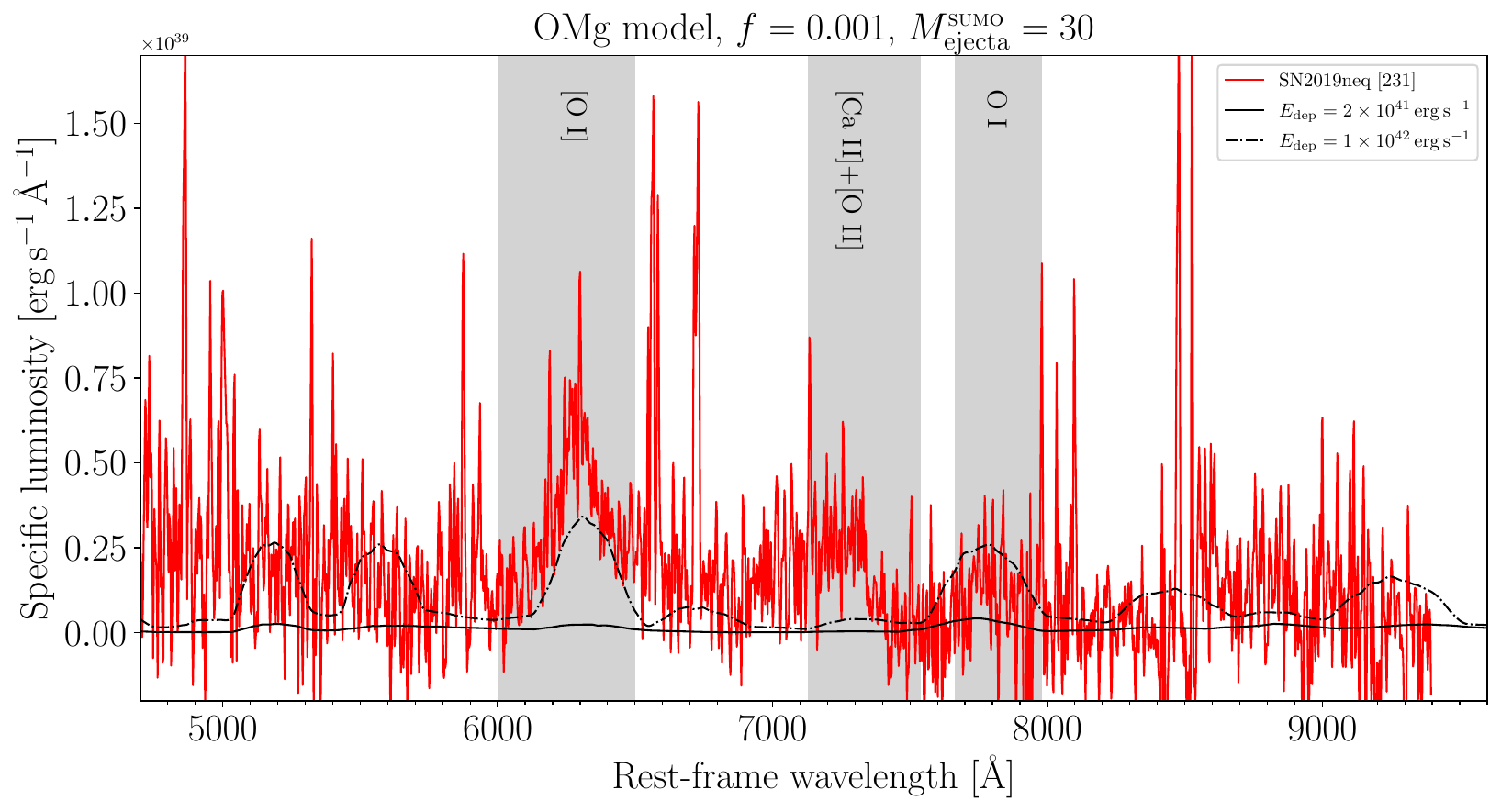}
    \caption{Similar to Fig.~\ref{fig:sumo1}, but for the OMg model and $M_{\rm ejecta}^\textsc{sumo}=30$. Here the solid line corresponds to an energy deposition $E_{\rm dep}=2\times10^{41}\,\mathrm{erg\,s^{-1}}$.}
    \label{fig:sumo5}
\end{figure*}
\begin{figure*}
    \centering
    \includegraphics[width=0.9\textwidth]{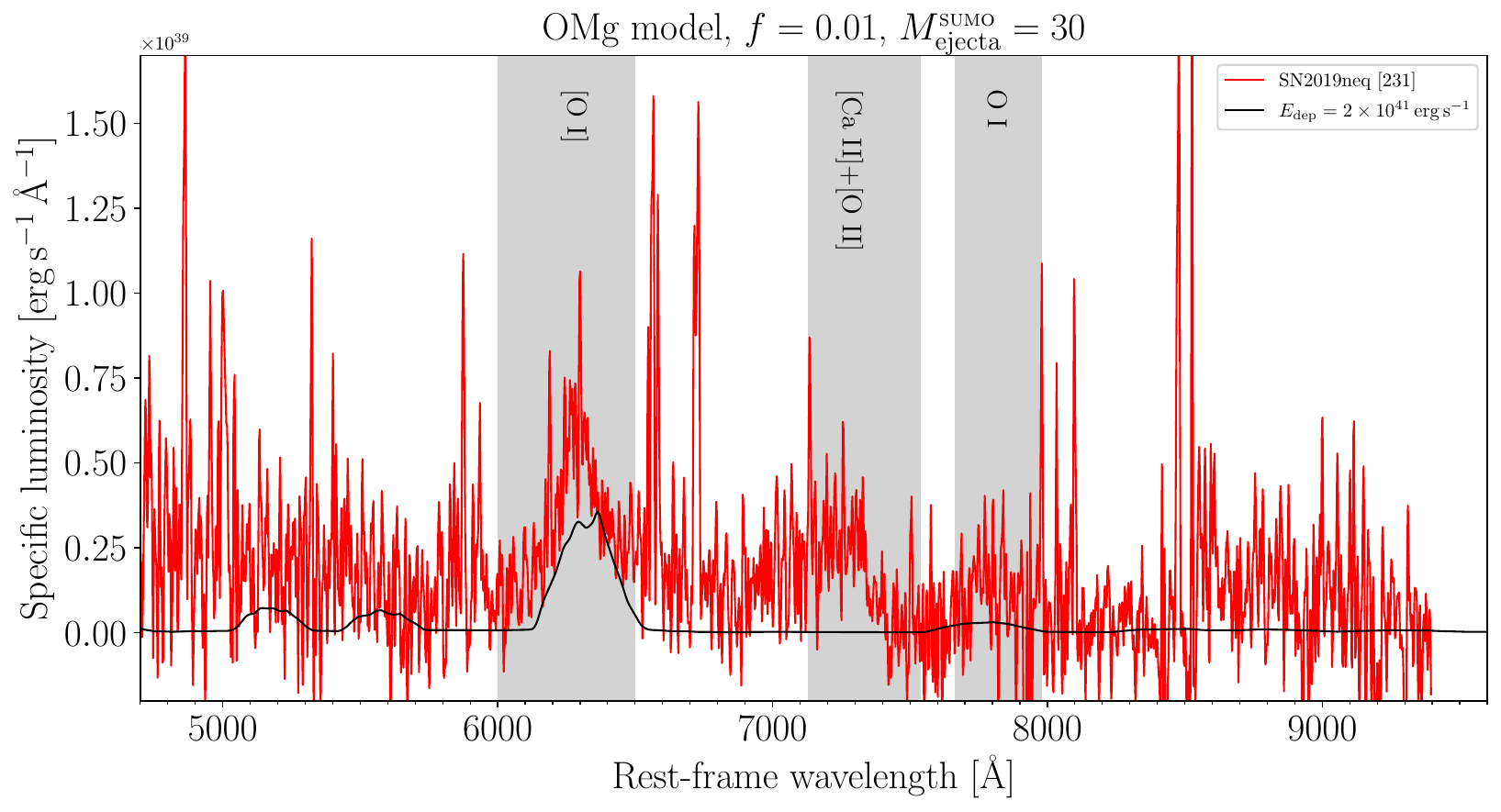}
    \caption{Similar to Fig.~\ref{fig:sumo1}, but for the OMg model, $f=0.01$ and $M_{\rm ejecta}^\textsc{sumo}=30$. Here the solid line corresponds to an energy deposition $E_{\rm dep}=2\times10^{41}\,\mathrm{erg\,s^{-1}}$.}
    \label{fig:sumo6}
\end{figure*}
\begin{figure*}
    \centering
    \includegraphics[width=0.9\textwidth]{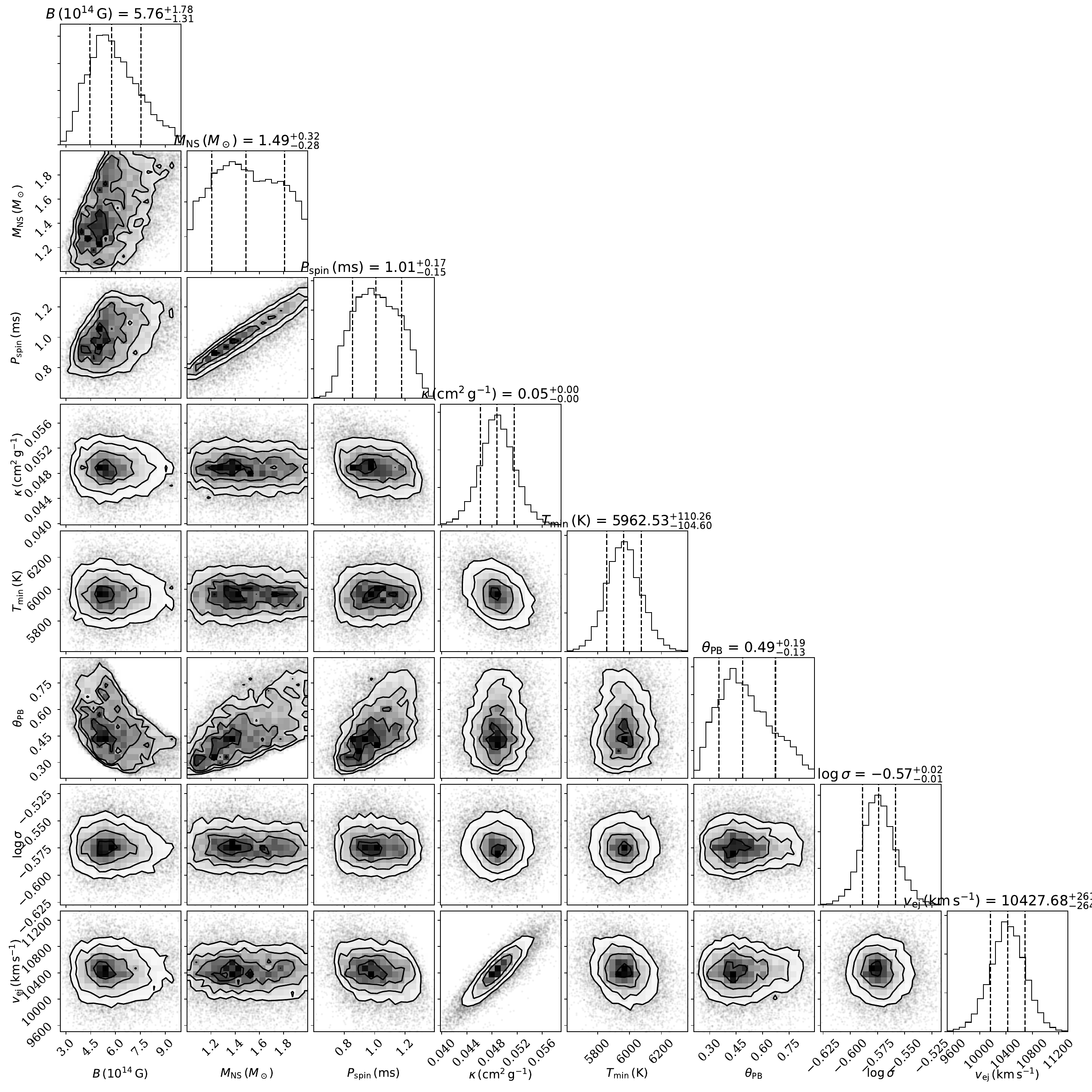}
    \caption{Corner plot with the best-fit parameters of the \textsc{mosfit} fit obtained with the \textsc{slsn} model.}
    \label{fig:corner_slsn}
\end{figure*}
\begin{figure*}
    \centering
    \includegraphics[width=0.9\textwidth]{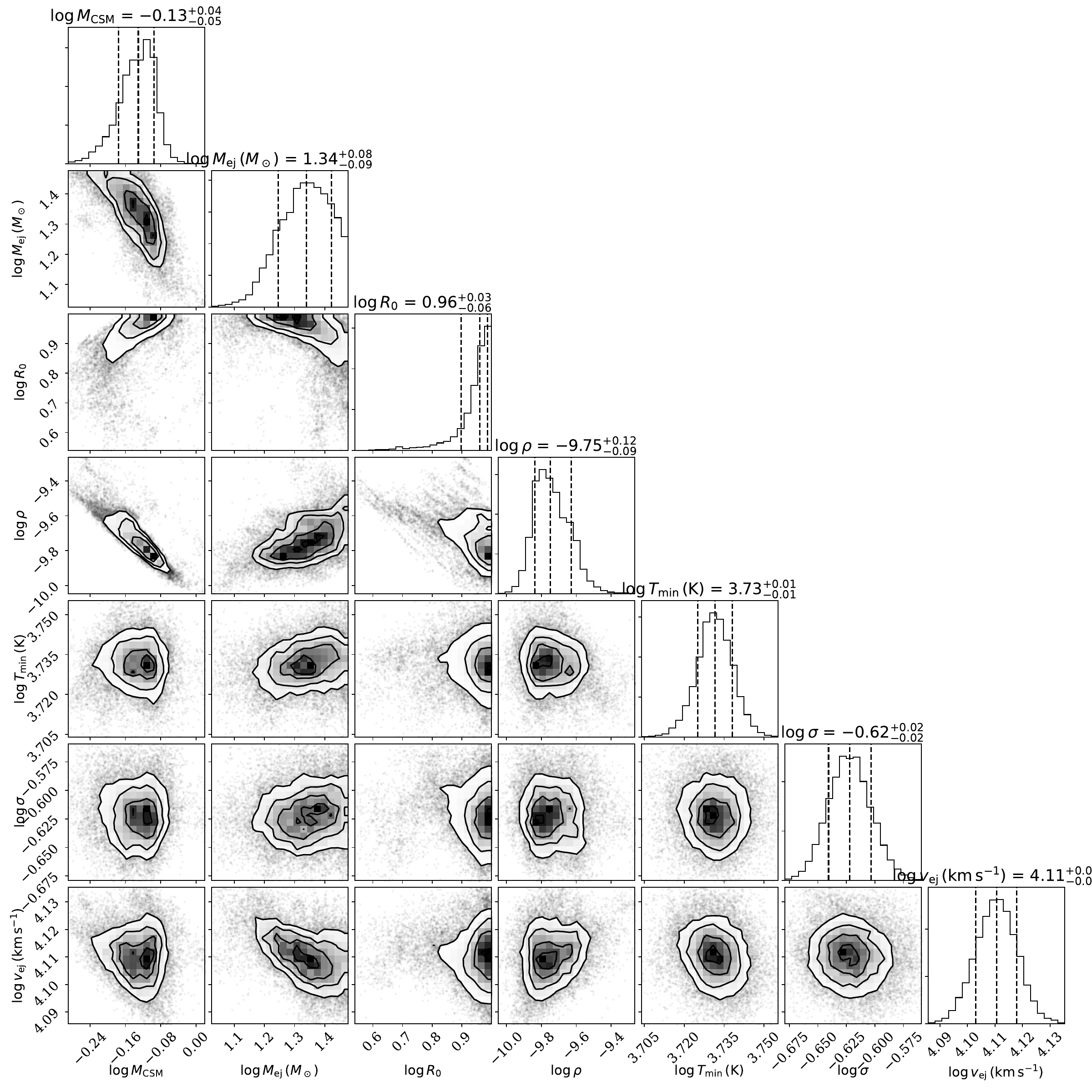}
    \caption{Same as Fig.~\ref{fig:corner_slsn}, but for the \textsc{csm} fit.}
    \label{fig:corner_csm}
\end{figure*}
\section{Tables}
\begin{table*}
\caption{$uvw1,uvm2,uvw2$-filters observed (non K-corrected, non S-corrected) {aperture} magnitudes (in AB system). Errors are in parentheses.}
\label{tab:19neq_uvottab}
\begin{tabular}{llllll}
\hline
MJD&phase&$uvw1$&$uvm2$&$uvw2$&instrument\\
&[days]&&&&\\
\hline
58729.66&-1.21&17.65(0.07)&18.26(0.07)&18.85(0.09)&\textit{Swift}/UVOT\\
58732.68&1.52&17.88(0.07)&18.44(0.07)&18.82(0.09)&\textit{Swift}/UVOT\\
58734.36&3.04&18.03(0.08)&18.52(0.08)&18.95(0.10)&\textit{Swift}/UVOT\\
58738.78&7.03&18.44(0.09)&18.93(0.09)&19.37(0.11)&\textit{Swift}/UVOT\\
58753.79&20.61&19.97(0.13)&20.49(0.12)&20.66(0.14)&\textit{Swift}/UVOT\\
58759.20&25.50&20.06(0.21)&20.55(0.18)&21.18(0.28)&\textit{Swift}/UVOT\\
58765.11&30.84&20.47(0.44)&20.95(0.32)&20.80(0.34)&\textit{Swift}/UVOT\\
58766.34&31.96&20.82(0.27)&21.34(0.21)&21.24(0.22)&\textit{Swift}/UVOT\\
\hline
\end{tabular}
\end{table*}
\begin{table*}
\caption{$u,g,r,i,z$-filter observed (non K-corrected, {non S-corrected}) magnitudes (in AB system). Errors are in parentheses.}
\label{tab:19neq_ugriztab}
\begin{tabular}{llllllll}
\hline
MJD&phase&$u$&$g$&$r$&$i$&$z$&instrument\\
&[days]&&&&&\\
\hline
58617.43&-102.69&-&$\gtrsim22.39$&-&-&-&ZTF\\
58644.22&-78.47&-&$\gtrsim22.36$&-&-&-&ZTF\\
58651.21&-72.15&-&$\gtrsim20.45$&-&-&-&ZTF\\
58658.44&-65.61&-&$\gtrsim21.78$&-&-&-&ZTF\\
58665.22&-59.48&-&$\gtrsim22.48$&-&-&-&ZTF\\
58679.30&-46.75&-&$\gtrsim21.58$&-&-&-&ZTF\\
58685.17&-41.44&-&$\gtrsim22.08$&-&-&-&ZTF\\
58692.23&-35.06&-&$\gtrsim21.82$&-&-&-&ZTF\\
58699.28&-29.95&-&-&$\gtrsim20.54$&-&-&ZTF\\
58699.31&-29.92&-&$\gtrsim20.72$&-&-&-&ZTF\\
58703.28&-26.17&-&-&$\gtrsim20.24$&-&-&ZTF\\
58703.31&-26.14&-&$\gtrsim20.71$&-&-&-&ZTF\\
58704.28&-24.16&-&$\gtrsim20.17$&-&-&-&ZTF\\
58704.31&-24.14&-&20.37(0.26)&-&-&-&ZTF\\
58704.39&-24.06&-&19.26(0.24)&-&-&-&ATLAS\\
58705.23&-23.31&-&19.78(0.24)&-&-&-&ZTF\\
58706.18&-22.45&-&19.54(0.21)&-&-&-&ZTF\\
58706.23&-22.40&-&-&-&-&-&ZTF\\
58707.23&-21.50&-&19.43(0.17)&-&-&-&ZTF\\
58708.20&-20.62&-&19.24(0.16)&19.54(0.17)&-&-&ZTF\\
58708.41&-20.43&-&18.57(0.30)&-&-&-&ATLAS\\
58709.20&-19.71&-&19.20(0.14)&19.29(0.14)&-&-&ZTF\\
58709.26&-19.66&-&19.12(0.13)&19.51(0.19)&-&-&ZTF\\
58710.16&-18.85&-&18.90(0.10)&-&-&-&ZTF\\
58710.33&-18.69&-&-&18.68(0.52)&-&-&ATLAS\\
58712.25&-16.96&-&18.47(0.07)&18.76(0.10)&-&-&ZTF\\
58713.24&-16.06&-&17.95(0.24)&18.49(0.10)&-&-&ZTF\\
58714.33&-15.07&-&-&18.52(0.37)&-&-&ATLAS\\
58714.33&-15.07&-&-&18.26(0.33)&-&-&ATLAS\\
58714.34&-15.06&-&-&18.36(0.35)&-&-&ATLAS\\
58714.35&-15.06&-&-&18.23(0.33)&-&-&ATLAS\\
58715.33&-14.17&-&17.91(0.06)&18.12(0.07)&-&-&ZTF\\
58716.37&-13.23&-&17.74(0.05)&17.98(0.08)&-&-&ZTF\\
58716.45&-13.16&-&-&18.14(0.32)&-&-&ATLAS\\
58716.45&-13.16&-&-&17.85(0.28)&-&-&ATLAS\\
58716.46&-13.15&-&-&18.05(0.32)&-&-&ATLAS\\
58716.47&-13.14&-&-&18.11(0.30)&-&-&ATLAS\\
58718.18&-11.59&-&17.52(0.04)&-&-&-&ZTF\\
58718.18&-11.59&-&17.52(0.04)&-&-&-&ZTF\\
58719.32&-10.56&-&17.43(0.06)&17.70(0.06)&-&-&ZTF\\
58719.32&-10.56&-&17.43(0.06)&17.70(0.06)&-&-&ZTF\\
58720.32&-9.66&-&17.39(0.03)&-&-&-&ATLAS\\
58720.32&-9.66&-&17.39(0.03)&-&-&-&ATLAS\\
58720.33&-9.65&-&17.41(0.03)&-&-&-&ATLAS\\
58720.34&-9.64&-&17.36(0.03)&-&-&-&ATLAS\\
58722.43&-7.75&-&-&17.36(0.20)&-&-&ATLAS\\
58722.44&-7.74&-&-&17.38(0.20)&-&-&ATLAS\\
58722.45&-7.73&-&-&17.32(0.20)&-&-&ATLAS\\
58723.18&-7.07&-&17.18(0.05)&-&-&-&ZTF\\
58723.18&-7.07&-&17.16(0.04)&-&-&-&ZTF\\
58724.34&-6.02&-&17.11(0.03)&-&-&-&ATLAS\\
58724.34&-6.02&-&17.09(0.02)&-&-&-&ATLAS\\
58724.35&-6.01&-&17.14(0.02)&-&-&-&ATLAS\\
58724.36&-6.00&-&17.15(0.03)&-&-&-&ATLAS\\
58724.36&-6.00&-&17.17(0.03)&-&-&-&ATLAS\\
58724.36&-6.00&-&17.17(0.03)&-&-&-&ATLAS\\
58724.37&-6.00&-&17.17(0.03)&-&-&-&ATLAS\\
58724.38&-5.99&-&17.23(0.04)&-&-&-&ATLAS\\
58726.29&-4.26&-&-&17.22(0.18)&-&-&ATLAS\\
58726.30&-4.25&-&-&17.29(0.18)&-&-&ATLAS\\
58726.30&-4.25&-&-&17.22(0.18)&-&-&ATLAS\\
58726.31&-4.24&-&-&17.17(0.25)&-&-&ATLAS\\
\hline
\end{tabular}
\end{table*}
\begin{table*}
\begin{tabular}{llllllll}
\hline
58727.13&-3.50&-&17.09(0.04)&17.30(0.06)&-&-&ZTF\\
58727.13&-3.50&-&17.08(0.05)&17.31(0.04)&-&-&ZTF\\
58728.27&-2.47&-&17.03(0.02)&-&-&-&ATLAS\\
58728.28&-2.46&-&17.06(0.02)&-&-&-&ATLAS\\
58728.28&-2.46&-&17.11(0.02)&-&-&-&ATLAS\\
58728.29&-2.45&-&17.04(0.02)&-&-&-&ATLAS\\
58728.30&-2.44&-&17.12(0.03)&-&-&-&ATLAS\\
58728.30&-2.44&-&17.06(0.02)&-&-&-&ATLAS\\
58728.31&-2.43&-&17.10(0.03)&-&-&-&ATLAS\\
58728.32&-2.42&-&17.26(0.22)&-&-&-&ATLAS\\
58728.53&-2.24&-&17.09(0.01)&17.36(0.01)&17.45(0.01)&-&TNT\\
58729.07&-1.75&17.17(0.04)&17.01(0.03)&17.28(0.03)&17.40(0.03)&17.63(0.03)&AFOSC\\
58729.56&-1.31&-&17.03(0.01)&17.30(0.01)&17.41(0.02)&-&TNT\\
58729.91&-0.99&17.35(0.04)&17.06(0.00)&17.20(0.02)&17.37(0.02)&17.55(0.01)&ALFOSC\\
58730.16&-0.76&-&-&17.27(0.05)&-&-&ZTF\\
58730.38&-0.56&-&-&17.20(0.18)&-&-&ATLAS\\
58730.39&-0.55&-&-&17.16(0.18)&-&-&ATLAS\\
58730.39&-0.55&-&-&17.26(0.20)&-&-&ATLAS\\
58730.40&-0.54&-&-&17.19(0.18)&-&-&ATLAS\\
58730.57&-0.39&-&17.07(0.01)&17.31(0.02)&17.36(0.02)&-&TNT\\
58730.80&-0.18&17.02(0.03)&17.11(0.01)&17.24(0.02)&17.43(0.02)&-&Schmidt\\
58731.15&0.14&-&17.05(0.23)&17.48(0.16)&-&-&ZTF\\
58731.57&0.52&-&17.04(0.01)&17.26(0.01)&17.39(0.01)&-&TNT\\
58732.33&1.20&-&-&17.26(0.18)&-&-&ATLAS\\
58732.33&1.20&-&-&17.20(0.18)&-&-&ATLAS\\
58732.34&1.21&-&-&17.16(0.20)&-&-&ATLAS\\
58732.35&1.22&-&-&17.18(0.20)&-&-&ATLAS\\
58732.56&1.41&-&17.04(0.01)&17.24(0.01)&17.35(0.01)&-&TNT\\
58733.24&2.02&-&17.08(0.04)&17.25(0.05)&-&-&ZTF\\
58733.57&2.32&-&17.01(0.02)&17.29(0.02)&17.35(0.01)&-&TNT\\
58734.14&2.84&-&17.09(0.05)&17.25(0.04)&-&-&ZTF\\
58734.27&2.96&-&-&17.21(0.18)&-&-&ATLAS\\
58734.28&2.97&-&-&17.14(0.18)&-&-&ATLAS\\
58734.28&2.97&-&-&17.20(0.18)&-&-&ATLAS\\
58734.29&2.97&-&-&17.13(0.18)&-&-&ATLAS\\
58734.57&3.22&-&17.01(0.02)&17.27(0.02)&17.32(0.02)&-&TNT\\
58734.88&3.50&17.46(0.02)&17.07(0.01)&17.22(0.01)&17.36(0.01)&17.55(0.02)&ALFOSC\\
58735.82&4.35&17.17(0.04)&17.11(0.02)&-&-&-&Schmidt\\
58735.82&4.35&17.20(0.04)&17.18(0.03)&-&-&-&Schmidt\\
58736.29&4.78&-&-&17.22(0.20)&-&-&ATLAS\\
58736.30&4.79&-&-&17.22(0.20)&-&-&ATLAS\\
58736.30&4.79&-&-&17.19(0.20)&-&-&ATLAS\\
58736.31&4.80&-&-&17.28(0.20)&-&-&ATLAS\\
58736.81&5.25&17.28(0.01)&17.23(0.02)&17.39(0.02)&17.50(0.02)&-&Schmidt\\
58736.82&5.26&17.29(0.02)&17.14(0.01)&17.30(0.01)&17.39(0.02)&-&Schmidt\\
58737.14&5.55&-&17.13(0.04)&-&-&-&ZTF\\
58737.14&5.55&-&17.15(0.04)&-&-&-&ZTF\\
58737.80&6.15&17.30(0.01)&17.35(0.03)&17.36(0.03)&17.50(0.03)&-&Schmidt\\
58737.80&6.15&17.28(0.02)&17.21(0.02)&17.36(0.03)&17.59(0.03)&-&Schmidt\\
58738.81&7.06&17.40(0.03)&17.24(0.04)&17.59(0.06)&17.42(0.04)&-&Schmidt\\
58738.81&7.06&17.42(0.03)&17.23(0.03)&17.37(0.03)&17.38(0.03)&-&Schmidt\\
58738.87&7.12&-&17.33(0.01)&17.45(0.01)&17.53(0.01)&17.49(0.02)&IO:O\\
58739.79&7.95&17.63(0.03)&17.32(0.02)&17.37(0.02)&17.42(0.03)&-&Schmidt\\
58739.88&8.03&-&17.30(0.02)&17.43(0.02)&17.43(0.02)&17.61(0.03)&IO:O\\
58740.26&8.38&-&17.23(0.05)&17.37(0.05)&-&-&ZTF\\
58740.27&8.38&-&17.25(0.04)&17.32(0.04)&-&-&ZTF\\
58740.35&8.45&-&-&17.26(0.23)&-&-&ATLAS\\
58740.36&8.46&-&-&17.26(0.25)&-&-&ATLAS\\
58740.36&8.46&-&-&17.22(0.25)&-&-&ATLAS\\
58740.37&8.47&-&-&17.25(0.25)&-&-&ATLAS\\
58740.89&8.94&-&17.41(0.01)&17.52(0.01)&17.54(0.01)&17.57(0.01)&IO:O\\
58741.86&9.82&17.95(0.04)&17.40(0.01)&17.38(0.02)&17.43(0.01)&17.61(0.01)&ALFOSC\\
\hline
\end{tabular}
\end{table*}
\begin{table*}
\begin{tabular}{llllllll}
\hline
58742.01&9.96&-&17.42(0.01)&17.44(0.01)&17.49(0.01)&18.02(0.05)&IO:O\\
58742.93&10.79&-&17.55(0.01)&17.52(0.01)&17.52(0.01)&17.67(0.02)&IO:O\\
58743.14&10.97&-&17.45(0.06)&-&-&-&ZTF\\
58743.54&11.33&-&17.66(0.03)&17.59(0.02)&17.58(0.02)&-&TNT\\
58744.14&11.88&-&17.49(0.05)&-&-&-&ZTF\\
58744.27&12.00&-&-&17.44(0.20)&-&-&ATLAS\\
58744.27&12.00&-&-&17.34(0.18)&-&-&ATLAS\\
58744.28&12.01&-&-&17.44(0.20)&-&-&ATLAS\\
58744.29&12.02&-&-&17.41(0.20)&-&-&ATLAS\\
58744.53&12.23&-&17.59(0.02)&17.56(0.02)&17.53(0.02)&-&TNT\\
58745.50&13.11&-&17.67(0.02)&17.58(0.02)&17.58(0.02)&-&TNT\\
58746.14&13.69&-&17.62(0.05)&-&-&-&ZTF\\
58746.31&13.84&-&-&17.48(0.25)&-&-&ATLAS\\
58746.31&13.84&-&-&17.36(0.25)&-&-&ATLAS\\
58746.32&13.85&-&-&17.45(0.23)&-&-&ATLAS\\
58746.33&13.86&-&-&17.51(0.20)&-&-&ATLAS\\
58747.13&14.59&-&17.70(0.05)&-&-&-&ZTF\\
58747.50&14.92&-&17.71(0.02)&17.61(0.02)&17.53(0.02)&-&TNT\\
58748.34&15.68&-&-&17.65(0.20)&-&-&ATLAS\\
58748.34&15.68&-&-&17.64(0.20)&-&-&ATLAS\\
58748.35&15.69&-&-&17.54(0.35)&-&-&ATLAS\\
58748.36&15.70&-&-&17.54(0.25)&-&-&ATLAS\\
58748.51&15.84&-&17.85(0.02)&17.68(0.01)&17.65(0.01)&-&TNT\\
58749.26&16.51&-&-&17.67(0.06)&-&-&ZTF\\
58749.54&16.76&-&17.89(0.01)&17.72(0.01)&17.68(0.01)&-&TNT\\
58750.14&17.31&-&17.92(0.06)&-&-&-&ZTF\\
58750.89&17.98&18.83(0.06)&18.04(0.02)&17.67(0.02)&17.64(0.04)&-&Schmidt\\
58750.89&17.98&18.80(0.06)&18.04(0.02)&17.67(0.02)&17.64(0.04)&-&Schmidt\\
58751.86&18.86&19.00(0.04)&18.05(0.01)&17.71(0.01)&17.65(0.02)&17.67(0.02)&ALFOSC\\
58752.25&19.21&-&18.22(0.08)&-&-&-&ATLAS\\
58752.25&19.21&-&18.00(0.05)&-&-&-&ATLAS\\
58752.26&19.22&-&18.02(0.05)&-&-&-&ATLAS\\
58752.28&19.24&-&18.00(0.04)&-&-&-&ATLAS\\
58754.27&21.04&-&18.18(0.07)&-&-&-&ZTF\\
58754.29&21.06&-&-&17.84(0.20)&-&-&ATLAS\\
58754.30&21.07&-&-&17.82(0.20)&-&-&ATLAS\\
58754.30&21.07&-&-&17.74(0.20)&-&-&ATLAS\\
58754.31&21.08&-&-&17.85(0.25)&-&-&ATLAS\\
58755.46&22.11&-&18.31(0.05)&17.98(0.03)&17.78(0.01)&-&TNT\\
58756.29&22.87&-&18.18(0.06)&-&-&-&ATLAS\\
58756.29&22.87&-&18.19(0.06)&-&-&-&ATLAS\\
58756.30&22.88&-&18.13(0.06)&-&-&-&ATLAS\\
58756.31&22.89&-&18.23(0.06)&-&-&-&ATLAS\\
58757.53&23.98&-&18.49(0.04)&17.98(0.03)&17.87(0.02)&-&TNT\\
58758.23&24.62&-&-&17.86(0.20)&-&-&ATLAS\\
58758.24&24.63&-&-&17.94(0.20)&-&-&ATLAS\\
58758.24&24.63&-&-&17.95(0.20)&-&-&ATLAS\\
58758.25&24.64&-&-&18.02(0.23)&-&-&ATLAS\\
58758.54&24.90&-&18.16(0.03)&17.77(0.03)&17.64(0.02)&-&TNT\\
58759.77&26.02&19.58(0.15)&18.52(0.03)&17.93(0.03)&17.79(0.03)&-&Schmidt\\
58759.89&26.12&19.59(0.04)&18.50(0.01)&18.00(0.01)&17.85(0.01)&17.90(0.01)&ALFOSC\\
58760.14&26.35&-&18.41(0.07)&-&-&-&ZTF\\
58762.27&28.28&-&-&17.87(0.25)&-&-&ATLAS\\
58762.28&28.28&-&-&17.85(0.25)&-&-&ATLAS\\
58762.28&28.28&-&-&17.76(0.23)&-&-&ATLAS\\
58762.29&28.29&-&-&17.90(0.25)&-&-&ATLAS\\
58764.12&29.95&-&-&18.02(0.06)&-&-&ZTF\\
58764.12&29.95&-&18.59(0.10)&18.00(0.07)&-&-&ZTF\\
58764.26&30.07&-&-&18.14(0.28)&-&-&ATLAS\\
58764.27&30.08&-&-&18.16(0.28)&-&-&ATLAS\\
58764.27&30.08&-&-&18.09(0.28)&-&-&ATLAS\\
58764.28&30.09&-&-&18.09(0.28)&-&-&ATLAS\\
58764.49&30.28&-&18.68(0.03)&18.17(0.02)&18.00(0.02)&-&TNT\\
58765.47&31.17&-&18.71(0.04)&18.10(0.08)&17.98(0.04)&-&TNT\\
58765.88&31.53&19.93(0.03)&18.75(0.01)&18.10(0.02)&17.99(0.01)&18.08(0.01)&ALFOSC\\
58765.94&31.59&-&18.87(0.01)&18.16(0.01)&18.08(0.01)&18.06(0.01)&IO:O\\
58767.12&32.66&-&-&18.13(0.08)&-&-&ZTF\\
\hline
\end{tabular}
\end{table*}
\begin{table*}
\begin{tabular}{llllllll}
\hline
58767.12&32.66&-&18.60(0.40)&18.09(0.09)&-&-&ZTF\\
58767.86&33.33&-&18.80(0.01)&18.25(0.01)&18.34(0.04)&18.11(0.02)&IO:O\\
58768.20&33.64&-&-&18.09(0.45)&-&-&ATLAS\\
58768.21&33.65&-&-&18.21(0.41)&-&-&ATLAS\\
58768.21&33.65&-&-&18.17(0.48)&-&-&ATLAS\\
58768.22&33.66&-&-&18.15(0.45)&-&-&ATLAS\\
58770.20&35.45&-&-&18.07(0.08)&-&-&ZTF\\
58770.20&35.45&-&18.92(0.14)&18.17(0.12)&-&-&ZTF\\
58770.26&35.50&-&-&18.31(0.39)&-&-&ATLAS\\
58770.26&35.50&-&-&18.09(0.32)&-&-&ATLAS\\
58770.26&35.50&-&-&18.52(0.45)&-&-&ATLAS\\
58770.26&35.50&-&-&18.20(0.33)&-&-&ATLAS\\
58770.27&35.51&-&-&18.12(0.36)&-&-&ATLAS\\
58770.27&35.51&-&-&18.25(0.35)&-&-&ATLAS\\
58770.28&35.52&-&-&18.25(0.36)&-&-&ATLAS\\
58770.28&35.52&-&-&18.19(0.33)&-&-&ATLAS\\
58771.92&37.00&-&19.01(0.04)&18.32(0.02)&18.17(0.01)&18.20(0.02)&IO:O\\
58772.27&37.32&-&-&18.45(0.36)&-&-&ATLAS\\
58772.27&37.32&-&-&18.43(0.36)&-&-&ATLAS\\
58772.28&37.33&-&-&18.17(0.35)&-&-&ATLAS\\
58772.29&37.34&-&-&18.18(0.35)&-&-&ATLAS\\
58774.20&39.06&-&-&18.31(0.33)&-&-&ATLAS\\
58774.20&39.06&-&-&18.78(0.36)&-&-&ATLAS\\
58774.20&39.06&-&-&18.30(0.28)&-&-&ATLAS\\
58774.21&39.07&-&-&18.31(0.25)&-&-&ATLAS\\
58774.21&39.07&-&-&18.54(0.27)&-&-&ATLAS\\
58774.22&39.08&-&-&18.33(0.25)&-&-&ATLAS\\
58776.53&41.16&-&19.27(0.04)&18.52(0.03)&18.33(0.03)&-&TNT\\
58777.50&42.04&-&19.29(0.04)&18.56(0.03)&18.31(0.03)&-&TNT\\
58778.27&42.74&-&-&18.50(0.30)&-&-&ATLAS\\
58778.28&42.75&-&-&18.93(0.35)&-&-&ATLAS\\
58778.28&42.75&-&-&18.61(0.30)&-&-&ATLAS\\
58778.30&42.77&-&-&18.46(0.32)&-&-&ATLAS\\
58778.80&43.22&20.56(0.19)&19.33(0.08)&18.58(0.02)&18.36(0.06)&-&Schmidt\\
58778.82&43.24&20.79(0.05)&19.33(0.01)&18.56(0.01)&18.39(0.01)&18.30(0.02)&ALFOSC\\
58779.76&44.09&-&19.35(0.08)&18.64(0.08)&18.40(0.09)&-&Schmidt\\
58780.25&44.53&-&19.30(0.12)&-&-&-&ATLAS\\
58780.26&44.54&-&19.17(0.10)&-&-&-&ATLAS\\
58780.26&44.54&-&19.53(0.13)&-&-&-&ATLAS\\
58780.27&44.55&-&18.97(0.09)&-&-&-&ATLAS\\
58782.50&46.56&-&19.79(0.09)&18.80(0.05)&18.63(0.05)&-&TNT\\
58783.75&47.70&21.04(0.12)&19.63(0.05)&18.88(0.03)&18.63(0.01)&18.55(0.07)&AFOSC\\
58788.22&51.74&-&-&19.19(0.39)&-&-&ATLAS\\
58788.22&51.74&-&-&19.43(0.42)&-&-&ATLAS\\
58788.23&51.75&-&-&19.42(0.44)&-&-&ATLAS\\
58788.24&51.76&-&-&19.53(0.47)&-&-&ATLAS\\
58793.46&56.47&-&20.32(0.24)&19.35(0.08)&19.27(0.07)&-&TNT\\
58802.20&64.38&-&-&19.88(0.52)&-&-&ATLAS\\
58802.21&64.39&-&-&19.70(0.48)&-&-&ATLAS\\
58802.22&64.40&-&-&19.64(0.47)&-&-&ATLAS\\
58805.81&67.64&-&20.56(0.02)&19.68(0.01)&19.38(0.01)&19.26(0.03)&ALFOSC\\
58806.43&68.20&-&20.72(0.15)&19.77(0.08)&19.38(0.07)&-&TNT\\
58814.43&75.43&-&20.82(0.17)&19.56(0.13)&-&-&TNT\\
58918.17&169.24&-&-&$\gtrsim$21.27&-&-&ALFOSC\\
58937.20&186.45&-&-&$\gtrsim$22.78&$\gtrsim$22.32&$\gtrsim$20.61&ALFOSC\\
58986.14&230.70&-&-&$\gtrsim$23.23&-&-&OSIRIS\\
\hline
\end{tabular}
\end{table*}
\begin{table*}
\centering
\caption{$U,B,V$-observed (non K-corrected, {non S-corrected}) magnitudes (in AB system). {\textit{Swift}/UVOT photometry was measured with a 5''-radius aperture (see text)}. Errors are in parentheses.}
\label{tab:19neq_bvtab}
\begin{tabular}{llllll}
\hline
MJD&phase&$U$&$B$&$V$&instrument\\
&[days]&&&&\\
\hline
58728.51&-2.25&-&17.27(0.01)&17.15(0.01)&TNT\\
58729.07&-1.75&-&17.14(0.03)&17.16(0.02)&AFOSC\\
58729.54&-1.32&-&17.17(0.01)&17.10(0.02)&TNT\\
58729.66&-1.21&16.99(0.07)&16.81(0.07)&17.03(0.11)&\textit{Swift}/UVOT\\
58729.91&-0.99&-&17.20(0.01)&17.13(0.02)&ALFOSC\\
58730.56&-0.40&-&17.14(0.02)&17.15(0.02)&TNT\\
58730.79&-0.19&-&17.08(0.01)&17.21(0.02)&Schmidt\\
58731.57&0.51&-&17.12(0.02)&17.09(0.01)&TNT\\
58732.55&1.40&-&17.15(0.01)&17.07(0.02)&TNT\\
58732.68&1.52&16.94(0.07)&17.08(0.07)&16.88(0.11)&\textit{Swift}/UVOT\\
58733.56&2.31&-&17.20(0.03)&17.06(0.02)&TNT\\
58734.36&3.04&16.99(0.07)&17.02(0.08)&16.88(0.12)&\textit{Swift}/UVOT\\
58734.56&3.22&-&17.15(0.02)&17.04(0.02)&TNT\\
58734.88&3.51&-&17.21(0.01)&17.18(0.01)&ALFOSC\\
58735.80&4.34&-&17.14(0.03)&-&Schmidt\\
58737.79&6.14&-&17.24(0.01)&17.59(0.02)&Schmidt\\
58737.79&6.14&-&17.16(0.01)&17.49(0.02)&Schmidt\\
58737.79&6.14&-&17.19(0.02)&17.25(0.02)&Schmidt\\
58738.78&7.03&17.32(0.08)&17.17(0.09)&16.91(0.11)&\textit{Swift}/UVOT\\
58738.79&7.04&-&17.35(0.06)&17.31(0.04)&Schmidt\\
58738.79&7.04&-&17.26(0.06)&17.29(0.04)&Schmidt\\
58738.79&7.04&-&17.23(0.03)&17.28(0.03)&Schmidt\\
58738.87&7.12&-&17.27(0.01)&17.34(0.02)&IO:O\\
58739.78&7.94&-&17.35(0.02)&17.24(0.02)&Schmidt\\
58739.78&7.94&-&17.35(0.02)&17.44(0.02)&Schmidt\\
58739.78&7.94&-&17.27(0.02)&17.38(0.02)&Schmidt\\
58739.78&7.94&-&17.24(0.02)&17.33(0.03)&Schmidt\\
58739.88&8.03&-&17.38(0.03)&17.33(0.02)&IO:O\\
58740.88&8.93&-&17.40(0.01)&17.36(0.03)&IO:O\\
58741.86&9.82&-&17.48(0.02)&17.41(0.01)&ALFOSC\\
58742.01&9.96&-&17.57(0.03)&17.37(0.02)&IO:O\\
58742.93&10.79&-&17.51(0.01)&17.48(0.02)&IO:O\\
58743.53&11.33&-&17.63(0.03)&17.48(0.03)&TNT\\
58744.51&12.22&-&17.70(0.02)&17.52(0.02)&TNT\\
58745.49&13.10&-&17.75(0.03)&17.58(0.02)&TNT\\
58747.49&14.91&-&17.87(0.03)&17.61(0.03)&TNT\\
58748.50&15.82&-&18.03(0.03)&17.72(0.02)&TNT\\
58749.53&16.76&-&18.06(0.02)&17.77(0.02)&TNT\\
58750.88&17.97&-&18.12(0.03)&17.81(0.02)&Schmidt\\
58751.86&18.86&-&18.25(0.01)&17.88(0.01)&ALFOSC\\
58753.79&20.61&18.69(0.11)&18.33(0.11)&17.85(0.13)&\textit{Swift}/UVOT\\
58755.44&22.10&-&18.70(0.19)&17.97(0.05)&TNT\\
58757.52&23.98&-&18.72(0.09)&18.13(0.04)&TNT\\
58759.20&25.50&18.97(0.20)&18.76(0.28)&18.23(0.31)&\textit{Swift}/UVOT\\
58759.78&26.02&-&18.65(0.04)&18.19(0.02)&Schmidt\\
58759.86&26.10&-&18.74(0.01)&18.19(0.01)&ALFOSC\\
58764.47&30.27&-&18.98(0.03)&18.38(0.03)&TNT\\
58765.11&30.84&19.36(0.44)&18.87(0.51)&-&\textit{Swift}/UVOT\\
58765.46&31.16&-&19.21(0.18)&18.56(0.10)&TNT\\
58765.87&31.53&-&19.06(0.01)&18.38(0.01)&ALFOSC\\
58765.93&31.59&-&19.10(0.03)&18.39(0.02)&IO:O\\
58766.34&31.96&19.30(0.19)&18.68(0.19)&18.24(0.21)&\textit{Swift}/UVOT\\
58767.85&33.32&-&19.14(0.03)&18.44(0.02)&IO:O\\
58771.92&37.00&-&19.42(0.07)&18.61(0.03)&IO:O\\
58776.51&41.15&-&19.57(0.07)&18.82(0.05)&TNT\\
58777.48&42.03&-&19.61(0.07)&18.83(0.05)&TNT\\
58778.79&43.21&-&19.55(0.04)&18.90(0.03)&Schmidt\\
58778.81&43.23&-&19.73(0.02)&18.79(0.01)&ALFOSC\\
58779.76&44.09&-&19.65(0.10)&18.98(0.06)&Schmidt\\
58782.49&46.56&-&19.83(0.15)&19.11(0.08)&TNT\\
58783.77&47.72&-&19.93(0.06)&19.20(0.02)&AFOSC\\
\hline
\end{tabular}
\end{table*}
\begin{table*}
\centering
\begin{tabular}{llllll}
\hline
58785.46&49.25&-&20.11(0.12)&19.20(0.07)&TNT\\
58786.45&50.14&-&20.01(0.10)&19.16(0.06)&TNT\\
58787.46&51.05&-&19.95(0.08)&19.36(0.05)&TNT\\
58793.44&56.46&-&20.04(0.23)&19.66(0.17)&TNT\\
58794.44&57.36&-&20.03(0.19)&19.66(0.14)&TNT\\
58805.42&67.29&-&-&19.44(0.15)&TNT\\
58805.80&67.64&-&20.67(0.04)&-&ALFOSC\\
58806.42&68.19&-&20.02(0.18)&20.01(0.12)&TNT\\
58813.43&74.54&-&-&-&TNT\\
58814.42&75.43&-&20.44(0.25)&20.12(0.17)&TNT\\
\hline
\end{tabular}
\end{table*}
\begin{table*}
\centering
\caption{NIR-observed (non K-corrected) $J,H,K$ magnitudes (in AB system). Errors are in parentheses.}
\label{tab:19neq_jhktab}
\begin{tabular}{llllll}
\hline
MJD&phase&$J$&$H$&$K_\mathrm{s}$&instrument\\
&[days]&&&&\\
\hline
58760.82&26.96&16.95(0.03)&-&18.86(0.06)&NOT+NOTCAM\\
58780.45&44.71&18.32(0.04)&18.81(0.04)&19.29(0.07)&NOT+NOTCAM\\
58803.81&65.84&18.81(0.07)&19.17(0.08)&19.19(0.09)&NOT+NOTCAM\\
58911.27&163.00&21.50(0.08)&-&-&NOT+NOTCAM\\
\hline
\end{tabular}
\end{table*}
\begin{table*}
\caption{S-corrections for Schmidt and AFOSC filters.}
\label{tab:scorrasiago}
\begin{tabular}{ccccccc}
\hline
MJD&$B$&$g$&$V$&$r$&$i$\\
\hline
58729.05&0.03&0.01&0.02&0.16&0.07\\
58748.49&-0.12&0.01&0.04&0.09&0.02\\
58724.91&0.02&0.00&0.02&0.13&0.06\\
58734.91&0.01&0.00&0.02&0.12&0.05\\
58741.87&0.00&0.00&0.03&0.13&0.04\\
58751.86&-0.04&0.00&0.05&0.11&0.03\\
58765.84&0.05&0.00&0.06&0.10&0.03\\
58772.82&-0.06&0.00&0.06&0.09&0.02\\
58805.82&-0.05&0.02&0.06&0.12&0.05\\
58819.81&-0.07&0.02&0.06&0.08&-0.01\\
58986.15&-1.01&0.31&-0.04&0.04&0.03\\
\hline
\end{tabular}
\end{table*}
\begin{table*}
\caption{S-corrections for ALFOSC.}
\label{tab:scorrnot}
\begin{tabular}{ccccccc}
\hline
MJD&$B$&$g$&$V$&$r$&$i$\\
\hline
58729.05&0.00&0.01&0.01&0.01&-0.01\\
58748.49&-0.08&-0.03&0.02&-0.01&-0.03\\
58724.91&0.00&0.01&0.02&-0.01&-0.05\\
58734.91&-0.00&0.00&0.02&-0.01&-0.04\\
58741.87&-0.01&0.00&0.02&0.00&-0.03\\
58751.86&-0.03&-0.01&0.00&-0.01&-0.03\\
58765.84&-0.03&-0.02&-0.01&-0.02&-0.01\\
58772.82&-0.03&-0.03&-0.02&-0.03&-0.02\\
58805.82&-0.03&-0.03&-0.02&-0.00&0.02\\
58819.81&-0.03&-0.04&-0.02&-0.02&-0.06\\
58986.15&-0.85&0.10&-0.07&0.01&-0.02\\
\hline
\end{tabular}
\end{table*}
\begin{table*}
\caption{S-corrections for IO:O filters.}
\label{tab:scorrlt}
\begin{tabular}{ccccccc}
\hline
MJD&$B$&$g$&$V$&$r$&$i$\\
\hline
58729.05&-0.01&0.00&0.02&0.00&-0.01\\
58748.49&-0.01&0.00&0.02&-0.01&-0.03\\
58724.91&-0.00&0.01&0.02&-0.00&-0.04\\
58734.91&-0.00&-0.01&0.02&-0.00&-0.03\\
58741.87&-0.00&-0.01&0.02&-0.00&-0.02\\
58751.86&-0.01&0.01&-0.00&-0.01&-0.02\\
58765.84&-0.02&0.03&-0.02&-0.02&-0.02\\
58772.82&-0.02&0.04&-0.03&-0.02&-0.02\\
58805.82&-0.02&0.07&-0.04&-0.00&0.02\\
58819.81&-0.02&0.08&-0.04&-0.01&-0.05\\
58986.15&-0.82&0.57&-0.12&0.01&-0.01\\
\hline
\end{tabular}
\end{table*}
\begin{table*}
\caption{S-corrections for \textit{Swift}/UVOT filters.}
\label{tab:scorrswift}
\begin{tabular}{ccccccc}
\hline
MJD&$B$&$V$\\
\hline
58729.05&-0.01&0.02\\
58748.49&-0.00&-0.03\\
58724.91&-0.01&0.02\\
58734.91&-0.00&0.01\\
58741.87&-0.01&-0.00\\
58751.86&-0.03&-0.03\\
58765.84&-0.04&-0.06\\
58772.82&-0.05&-0.07\\
58805.82&-0.04&-0.05\\
58819.81&-0.05&-0.06\\
58986.15&-7.08&0.11\\
\hline
\end{tabular}
\end{table*}
\begin{table*}
    \centering
        \caption{Estimated uncertainties $\Delta S_{\rm corr}$ for the filters $u,U,z,J,H,K_{\rm s}$ (for each instrument) divided in two temperature ranges.}
        \label{tab:scorrerr}
    \begin{tabular}{lll}
    
    \hline
         &{$5000\,\mathrm{K}<T<8000\,\mathrm{K}$}&{$8000\,\mathrm{K}<T<12000\,\mathrm{K}$}  \\
         \hline
         {NOT+ALFOSC/NOTCam}&\begin{tabular}{l}{$\Delta S_{{\rm corr,}u}=0.05$}\\{$\Delta S_{{\rm corr,}z}=0.01$}\\{$\Delta S_{{\rm corr,}J}=0.1$}\\{$\Delta S_{{\rm corr,}H}=0.1$}\\{$\Delta S_{{\rm corr,}K_{\rm s}}=0.1$}\end{tabular}&\begin{tabular}{l}{$\Delta S_{{\rm corr,}u}=0.15$}\\{$\Delta S_{{\rm corr,}z}=0.02$}\\{$\Delta S_{{\rm corr,}J}=0.1$}\\{$\Delta S_{{\rm corr,}H}=0.1$}\\{$\Delta S_{{\rm corr,}K_{\rm s}}=0.1$}\end{tabular}\\
         \hline
         {Schmidt/AFOSC}&\begin{tabular}{l}{$\Delta S_{{\rm corr,}u}=0.04$}\\{$\Delta S_{{\rm corr,}z}=-0.02$}\end{tabular}&\begin{tabular}{l}{$\Delta S_{{\rm corr,}u}=0.15$}\\{$\Delta S_{{\rm corr,}z}=-0.01$}\end{tabular}\\
         \hline
         {LT+IO:O}&\begin{tabular}{l}{$\Delta S_{{\rm corr,}z}=0.02$}\end{tabular}&\begin{tabular}{l}{$\Delta S_{{\rm corr,}z}=0.01$}\end{tabular}\\
         \hline
         {\textit{Swift}/UVOT}&\begin{tabular}{l}{$\Delta S_{{\rm corr,}U}=-0.03$}\end{tabular}&\begin{tabular}{l}{$\Delta S_{{\rm corr,}U}=-0.16$}\end{tabular}\\
         \hline
    \end{tabular}
    \label{tab:deltascorr}
\end{table*}
\begin{table*}
\caption{K-corrections expressed in magnitudes.}
\label{tab:19neq_kcorr}
\begin{tabular}{lllllllllllllll}
\hline
phase&$uvw2$&$uvm2$&$uvw1$&$u$&$U$&$B$&$g$&$V$&$r$&$i$&$z$&$J$&$H$&$K_\mathrm{s}$\\
maximum$[\mathrm{days}]$&filter&filter&filter&filter&filter&filter&filter&filter&filter&filter&filter&filter&filter&filter\\
\hline
-6.07&-0.12&-0.21&-0.22&0.01&-0.15&-0.15&-0.10&-0.15&-0.10&-0.19&0.22&0.25&0.28&0.29\\
-1.95&-0.16&-0.26&-0.26&-0.03&-0.15&-0.15&-0.10&-0.15&-0.10&-0.19&0.21&0.25&0.27&0.29\\
3.92&-0.16&-0.26&-0.26&-0.03&-0.11&-0.11&-0.09&-0.11&-0.09&-0.17&0.21&0.25&0.27&0.29\\
10.88&-0.22&-0.33&-0.29&-0.09&-0.10&-0.10&-0.08&-0.10&-0.08&-0.17&0.19&0.23&0.26&0.28\\
20.87&-0.31&-0.34&-0.14&-0.27&-0.07&-0.07&-0.08&-0.07&-0.08&-0.16&0.13&0.19&0.23&0.26\\
34.86&-0.25&-0.15&0.02&-0.38&-0.05&-0.06&-0.07&-0.06&-0.07&-0.16&0.09&0.16&0.21&0.25\\
41.83&-0.24&-0.11&0.04&-0.40&-0.01&-0.03&-0.06&-0.03&-0.06&-0.14&0.08&0.16&0.21&0.24\\
74.84&-0.08&0.10&0.15&-0.53&0.12&0.09&-0.03&0.09&-0.03&-0.10&0.03&0.13&0.19&0.23\\
88.81&-0.09&0.09&0.15&-0.52&0.12&0.09&-0.03&0.09&-0.03&-0.10&0.03&0.13&0.19&0.23\\
\hline
\end{tabular}
\end{table*}
\begin{table*}
\centering
\caption{Logarithm of the bolometric luminosity of SN~2019neq integrated over the $uvw2,uvm2,uvw1,U,B,g,V,r,i,z,J,H,K_\mathrm{s}$ filters.}
\label{tab:19neq_blc}
\begin{tabular}{ll}
\hline
phase&$\log_{10}L_\mathrm{bol}$\\
$[\mathrm{days}]$&\\
\hline
-22.80&43.35(0.02)\\
-22.80&43.35(0.02)\\
-21.80&43.45(0.02)\\
-21.80&43.45(0.02)\\
-21.74&43.37(0.02)\\
-21.74&43.37(0.02)\\
-20.67&43.68(0.03)\\
-18.75&43.66(0.02)\\
-18.75&43.66(0.02)\\
-17.76&43.78(0.02)\\
-17.76&43.78(0.02)\\
-16.67&43.87(0.03)\\
-16.67&43.77(0.03)\\
-16.66&43.83(0.03)\\
-16.65&43.88(0.03)\\
-15.67&43.92(0.02)\\
-15.67&43.92(0.02)\\
-14.63&43.98(0.02)\\
-14.63&43.98(0.02)\\
-14.55&43.91(0.02)\\
-14.55&44.02(0.02)\\
-14.54&43.95(0.02)\\
-14.53&43.93(0.02)\\
-11.67&44.09(0.02)\\
-11.67&44.09(0.02)\\
-8.57&44.22(0.02)\\
-8.56&44.21(0.02)\\
-8.55&44.23(0.02)\\
-4.71&44.27(0.02)\\
-4.70&44.25(0.02)\\
-4.70&44.27(0.02)\\
-4.69&44.29(0.02)\\
-3.87&44.24(0.02)\\
-3.87&44.24(0.01)\\
-2.47&44.22(0.01)\\
-1.93&44.25(0.01)\\
-1.44&44.25(0.01)\\
-1.09&44.27(0.01)\\
-0.84&44.26(0.01)\\
-0.62&44.27(0.02)\\
-0.61&44.27(0.02)\\
-0.61&44.27(0.02)\\
-0.60&44.27(0.02)\\
-0.43&44.26(0.01)\\
-0.20&44.27(0.01)\\
0.16&44.25(0.02)\\
0.58&44.26(0.01)\\
1.33&44.26(0.02)\\
1.33&44.26(0.02)\\
1.34&44.27(0.02)\\
1.35&44.26(0.02)\\
1.56&44.26(0.01)\\
2.24&44.25(0.02)\\
2.57&44.24(0.01)\\
3.14&44.24(0.02)\\
3.27&44.24(0.02)\\
3.28&44.25(0.02)\\
3.28&44.24(0.02)\\
3.29&44.25(0.02)\\
3.57&44.23(0.01)\\
3.88&44.22(0.01)\\
\hline
\end{tabular}
\end{table*}
\begin{table*}
\centering
\begin{tabular}{ll}
\hline
5.29&44.20(0.02)\\
5.30&44.20(0.02)\\
5.30&44.21(0.02)\\
5.31&44.20(0.02)\\
5.81&44.18(0.01)\\
5.82&44.19(0.01)\\
6.80&44.16(0.01)\\
6.80&44.16(0.01)\\
7.81&44.15(0.02)\\
7.81&44.13(0.02)\\
7.87&44.14(0.01)\\
8.80&44.12(0.01)\\
8.88&44.11(0.01)\\
9.27&44.11(0.02)\\
9.27&44.12(0.02)\\
9.35&44.12(0.02)\\
9.36&44.13(0.02)\\
9.36&44.12(0.02)\\
9.37&44.12(0.02)\\
9.89&44.08(0.01)\\
10.87&44.08(0.01)\\
11.01&44.05(0.01)\\
11.94&44.04(0.01)\\
12.54&44.01(0.01)\\
13.27&44.02(0.02)\\
13.27&44.03(0.02)\\
13.28&44.02(0.02)\\
13.29&44.02(0.02)\\
13.53&44.00(0.01)\\
14.50&43.98(0.01)\\
15.31&43.98(0.02)\\
15.31&43.99(0.03)\\
15.32&43.98(0.02)\\
15.33&43.98(0.02)\\
16.50&43.95(0.01)\\
17.34&43.92(0.02)\\
17.34&43.92(0.02)\\
17.35&43.93(0.03)\\
17.36&43.93(0.02)\\
17.52&43.91(0.01)\\
18.26&43.90(0.02)\\
18.54&43.89(0.01)\\
19.89&43.88(0.01)\\
19.89&43.88(0.01)\\
20.86&43.85(0.01)\\
23.29&43.80(0.03)\\
23.30&43.81(0.03)\\
23.30&43.82(0.03)\\
23.31&43.80(0.03)\\
24.46&43.77(0.02)\\
26.53&43.74(0.02)\\
27.23&43.76(0.03)\\
27.24&43.75(0.03)\\
27.24&43.75(0.03)\\
27.25&43.74(0.03)\\
27.54&43.78(0.03)\\
28.77&43.73(0.02)\\
28.89&43.72(0.02)\\
31.27&43.72(0.04)\\
31.28&43.72(0.04)\\
31.28&43.73(0.04)\\
31.29&43.71(0.04)\\
33.12&43.68(0.03)\\
33.12&43.67(0.03)\\
\hline
\end{tabular}
\end{table*}
\begin{table*}
\centering
\begin{tabular}{ll}
\hline
33.26&43.66(0.04)\\
33.27&43.65(0.04)\\
33.27&43.66(0.04)\\
33.28&43.66(0.04)\\
33.49&43.65(0.03)\\
34.48&43.64(0.03)\\
34.88&43.64(0.02)\\
34.94&43.62(0.02)\\
36.12&43.63(0.03)\\
36.12&43.64(0.03)\\
36.86&43.59(0.02)\\
37.20&43.62(0.04)\\
37.21&43.60(0.04)\\
37.21&43.59(0.04)\\
37.22&43.60(0.04)\\
39.20&43.62(0.03)\\
39.20&43.59(0.03)\\
39.26&43.56(0.04)\\
39.26&43.61(0.04)\\
39.26&43.52(0.04)\\
39.26&43.59(0.04)\\
39.27&43.60(0.04)\\
39.27&43.58(0.04)\\
39.28&43.58(0.04)\\
39.28&43.59(0.04)\\
40.92&43.55(0.02)\\
41.27&43.53(0.03)\\
41.27&43.53(0.03)\\
41.28&43.58(0.04)\\
41.29&43.58(0.04)\\
43.20&43.54(0.04)\\
43.20&43.45(0.03)\\
43.20&43.54(0.03)\\
43.21&43.54(0.03)\\
43.21&43.49(0.03)\\
43.22&43.54(0.03)\\
45.53&43.48(0.02)\\
46.50&43.47(0.02)\\
47.27&43.48(0.03)\\
47.28&43.40(0.03)\\
47.28&43.46(0.03)\\
47.30&43.49(0.04)\\
47.80&43.46(0.02)\\
47.82&43.46(0.02)\\
48.76&43.44(0.03)\\
51.50&43.36(0.03)\\
52.76&43.35(0.02)\\
57.22&43.24(0.04)\\
57.22&43.27(0.04)\\
57.23&43.25(0.04)\\
57.24&43.24(0.04)\\
62.46&43.20(0.03)\\
71.20&43.10(0.04)\\
71.21&43.12(0.04)\\
71.22&43.12(0.04)\\
74.81&43.09(0.02)\\
75.43&43.08(0.03)\\
\hline
\end{tabular}
\end{table*}
\begin{table*}
\centering
\caption{Spectra in Fig.~\ref{fig:19neq_spec}.}
\label{tab:19neq_sfo}
\begin{tabular}{lllll}
\hline
MJD&phase&instrument&resolution\\
&[days]&&[\AA{}]\\
\hline
58724.0&-6&ALFOSC [gr4]&15\\
58729.0&-2&AFOSC [VPH6+VPH7]&18\\
58734.9&4&ALFOSC [gr4]&19\\
58741.9&10&ALFOSC [gr4]&15\\
58748.5&16&BFOSC [G4]&\\ 	
58751.9&19&ALFOSC [gr4]&18\\
58765.9&32&ALFOSC [gr4]&15\\
58772.8&38&ALFOSC [gr4]&15\\
58805.8&68&ALFOSC [gr4]&25\\
58819.8&80&OSIRIS [R1000I]&8\\
58986.1&231&OSIRIS [R1000R]&9\\
\hline
\end{tabular}
\end{table*}

\bsp	
\label{lastpage}
\end{document}